\documentclass[a4paper,fleqn,usenatbib]{mnras}

\usepackage{newtxtext,newtxmath}

\usepackage[T1]{fontenc}
\usepackage{ae,aecompl}

\usepackage{bm}
\usepackage[linesnumbered]{algorithm2e}
\usepackage{color,hyperref}
\usepackage{float}
\usepackage{amsmath}	
\usepackage{amssymb}	
\usepackage{graphicx}	

\definecolor{linkcolor}{rgb}{0,0,0.25}
\hypersetup{
  colorlinks=true,        
  linkcolor=linkcolor,    
  citecolor=linkcolor,    
  filecolor=linkcolor,    
  urlcolor=linkcolor      
}

\title[Blind chemical tagging with DBSCAN]{Blind chemical tagging with DBSCAN: prospects for spectroscopic surveys}

\author[Price-Jones \& Bovy]{
Natalie Price-Jones$^{1,2}$\thanks{E-mail: price-jones@astro.utoronto.ca}, \&
Jo Bovy$^{1,2,3}$
\\
$^{1}$Department of Astronomy and Astrophysics, University of Toronto, 50 St. George Street, Toronto ON M5S 3H4, Canada\\
$^{2}$ Dunlap Institute for Astronomy and Astrophysics, University of Toronto, 50 St. George Street, Toronto, ON M5S 3H4, Canada\\
$^{3}$ Alfred P. Sloan Fellow  
}

\date{Accepted XXX. Received YYY; in original form ZZZ}

\pubyear{2019}

\begin{document}
\label{firstpage}
\pagerange{\pageref{firstpage}--\pageref{lastpage}}
\maketitle

\begin{abstract}
Chemical tagging has great promise as a technique to unveil our Galaxy's history. Grouping stars based on their similar chemistry can establish details of the star formation and merger history of the Milky Way. With precise measurements of stellar chemistry, chemical tagging may be able to group together stars born from the same gas cloud, regardless of their current positions and kinematics. Successfully tagging these birth clusters requires high quality chemical space information and a good cluster-finding algorithm. To test the feasibility of chemical tagging on data from current and upcoming spectroscopic surveys, we construct a realistic set of synthetic clusters, creating both observed spectra and derived chemical abundances for each star. We use Density-Based Spatial Clustering of Applications with Noise (DBSCAN) to group stars based on their spectra or abundances; these groups are matched to input clusters and are found to be highly homogeneous and complete. The percentage of clusters with more than 10 members recovered is 40\% when tagging on abundances with uncertainties achievable with current techniques. Based on our fiducial model for the Milky Way, we predict recovering over 600 clusters with at least 10 observed members and 70\% membership homogeneity in a sample similar to the APOGEE survey. Tagging larger surveys like the GALAH survey and the future Milky Way Mapper in SDSS V could recover tens of thousands of clusters at high homogeneity. Access to so many unique co-eval clusters will transform how we understand the star formation history and chemical evolution of our Galaxy.
\end{abstract}

\begin{keywords}
methods: data analysis -- stars: abundances -- stars: statistics -- open clusters and associations: general
\end{keywords}

\section{Introduction}
\label{sec:intro}

While the largest surveys of our Galaxy remain primarily photometric and kinematic (e.g. Gaia - \citealt{GaiaCollaboration2016}), there are growing collections of spectroscopic data for hundreds of thousands of stars (e.g. RAVE - \citealt{Steinmetz2006}; Gaia-ESO \citealt{Gilmore2012}; LAMOST - \citealt{Zhao2012}; APOGEE - \citealt{Majewski2017}; GALAH - \citealt{DeSilva2015}). This spectroscopic information facilitates a chemical tagging approach to understanding our Galaxy's evolution. Chemical tagging is the process of using stars' chemical compositions to classify them into groups that share similar chemistry \citep{Freeman2002}. Studying chemically identified groupings of stars offers advantages over groups found based on shared kinematics, namely that stars with similar chemistry should share some details of their respective origins. Stars formed in the same giant molecular cloud are expected to disperse from their birth cluster in less than 100 Myr \citep{Lada2003} through random interactions with their environment, making it difficult to track their origins through their present day kinematics. However, a star's chemical composition will change in minor and predictable ways as it ages (e.g. \citealt{Kraft1994}, \citealt{Weiss2000}). Processes that regulate the observed surface composition are internal to the star and largely deterministic except in unusual situations like mass transfer from a companion. These deterministic processes, which include internal dredge-up (\citealt{Masseron2015}) and atomic diffusion (\citealt{Dotter2017}) can be modelled as a function of stellar mass and age. If stars share a common chemical origin, their dispersal in phase space is no obstacle to confirming their similarity in chemical space. 

In the weak limit of chemical tagging, chemically identified groups of stars are broader, and may be associated with previously identified kinematic structures like the thin disk, the thick disk, or the halo (e.g., \citealt{Hawkins2015}, \citealt{Wojno2016}). Chemical tagging in this limit also offers a way to explore substructure within broad kinematically identified populations (e.g., \citealt{Martell2010}, \citealt{Bensby2013}, \citealt{Schiavon2015}, \citealt{Recio-Blanco2017}). Blind approaches to chemical tagging in this limit seek to define broad categories of stars (e.g. \citealt{Hayes2018}), or reconstruct chemical evolutionary history (e.g. \citealt{Jofre2016}). Chemical tagging has also been successful in comparing populations with distinct star formation histories, such as globular clusters (e.g., \citealt{Schiavon2016}, \citealt{Tang2017}) and Milky Way satellites like the Sgr dwarf \citep{Hasselquist2017}. 

The stronger limit of chemical tagging, identifying clusters of stars born in the same gas cloud, is predicated on the assumption that stars that belong to the same birth cluster will be chemically similar to each other and distinct from those born in other gas clouds. Simulations indicate that turbulence in a giant molecular cloud should be sufficient to ensure that star forming gas is chemically well mixed throughout the entire cloud \citep{Feng2014}. Numerous studies have demonstrated that open clusters, a good present day proxy for undispersed birth clusters, are in fact chemically homogeneous to the limit of our ability to measure their abundances (e.g. \citealt{DeSilva2006}, \citealt{DeSilva2007}, \citealt{Bovy2016}). \citet{Liu2016} showed in their work that it is possible to find pairs of stars within the same open cluster that have distinct chemical signatures (see also \citealt{Liu2016a}). However, such clusters will still appear as a chemical space over-density as long as the intrinsic spread of the abundances within members of the cluster is lower than the measurement uncertainties. Even if this is not the case, clusters can still be recovered as chemical space over-densities if their chemical signatures are sufficiently unique.

In addition to birth cluster homogeneity, chemical tagging in the strong limit also requires that birth clusters be chemically distinct from each other. The chemical uniqueness of open clusters is still being explored (e.g \citealt{Blanco-Cuaresma2015}, \citealt{Price-Jones2018}), but it seems that sampling multiple non-correlated elements is sufficient to distinguish between open clusters \citep{Blanco-Cuaresma2015}. If the processes that modify surface chemical composition are well understood, chemically tagging stars in a sufficiently resolved chemical space would offer a way to group stars from the same birth cluster. This approach can be validated on open clusters but is even more powerful when applied to a large sample of field stars. Judicious application of strong chemical tagging to such a sample could identify members of dispersed birth clusters. This would not only constrain how and when birth clusters are dispersed, but also allow age constraints for all of the member stars (\citealt{Bland-Hawthorn2010}, \citealt{Mitschang2014}).

The prospects for chemical tagging to recover birth clusters have not been universally promising (e.g. \citealt{Mitschang2014}, \citealt{Ting2015a}, \citealt{Blanco-Cuaresma2016}). Some studies have argued that chemical spaces are limited in their ability to accurately distinguish clusters, and the presence of chemical doppelgangers among field stars presents an additional challenge \citep{Ness2018}. However, some attempts at blind chemical tagging (e.g. \citealt{Hogg2016}, \citealt{Blanco-Cuaresma2018}, \citealt{Chen2018}) have produced encouraging results, recovering known clusters as well as identifying new chemical space structure.

Chemical tagging requires high resolution chemical information in order to be effective. Previous work has focused on achieving this by improving the precision with which elemental abundances are measured. Traditional abundance derivation from model spectra fitting suffers the limitations of the models, which are necessarily simplified (\citealt{Smiljanic2014}, \citealt{Perez2015}). This often results in correlations between the derived abundances and other stellar properties that can affect the spectrum (e.g. \citealt{Holtzman2015}, \citealt{Jofre2018}, \citealt{Nissen2018}). Intrinsic correlations between different abundances mean that increasing the number of elements in chemical space may not offer increased leverage in an attempt to differentiate stars. It is therefore crucial to reduce uncertainties in this space in order to distinguish the chemical signatures of different clusters.  There have been many recent approaches to improving abundance derivation using data-driven methods to achieve higher precision if not accuracy (e.g, \citealt{Ness2015}, \citealt{Casey2016a}, \citealt{Rix2016}, \citealt{Leung2019}).  

One way to avoid the uncertainty inherent to derived abundances is to make direct use of the spectra. In \citet{Price-Jones2018}, we showed that high resolution spectra can be reduced through expectation-maximized principal component analysis (\citealt{Dempster1977}, \citealt{Roweis1997}) to approximately ten significant dimensions. The spectra in their raw form do include non-chemical information about the stellar photosphere, but with appropriate pre-processing may serve as a useful chemical space in which to identify structure.

If the observed chemical space is sufficiently well resolved, a clustering algorithm can be used to identify groups within that space. Previous work has made use of the k-means algorithm (e.g. \citealt{Hogg2016}), but this approach requires an estimate of the number of clusters present in the data. In this work we use density-based spatial clustering of applications with noise (DBSCAN - \citealt{Ester1996}) to identify groups of stars in a synthetic set of birth clusters. This algorithm has a wide variety of applications to Milky Way data, used to classify stars based on photometry (e.g. \citealt{Kaderali2019}) and spectra (e.g. \citealt{Traven2017}). It has also been employed in the weak limit of chemical tagging in \citet{Hayes2018} to divide stars into two major populations in low-metallicity abundance space. In the strong limit, DBSCAN has been compared to other clustering algorithms on a sample of stellar abundances collected from known open clusters in \citet{Blanco-Cuaresma2015}. An approach closer to our interest is \citet{Chen2018}, who use DBSCAN (as well as other clustering algorithms) to recover members of globular clusters from a large sample of field stars. The more complex chemical variation within globular clusters requires one to consider radial velocities in addition to their initial chemical space in order to successfully recover the globular clusters. 

In this work, we are interested in pure chemical tagging, and so use only the spectra or derived abundances of the stars in synthetic birth clusters as separate ways to probe chemical space. DBSCAN makes use of density-based analysis to internally determine the number of clusters present in the data and can flag stars as noise stars if they cannot be assigned to a sufficiently dense cluster. These properties are both  relevant when chemically tagging real data, where the true number of clusters will be unknown and the intrinsic cluster mass function will likely populate chemical space with many stars that are the only representatives of their cluster. Such stars will create a chemical space background from which larger clusters must be identified as over-densities, an excellent use case for DBSCAN. By testing on a synthetic dataset in this work we quantitatively assess how well the groups recovered by DBSCAN match the clusters we create to serve as input data. Our variation of the chemical space spread between members of the same cluster allows us to predict the parameterization of DBCSAN when applied to observations from different large spectroscopic surveys. 

In Section~\ref{sec:synthetic}, we explain how we generate our synthetic datasets, followed by a description of DBSCAN in Section~\ref{sec:identify}. Section~\ref{sec:results} describes the properties of clusters identified by DBSCAN in the synthetic data, and in Section~\ref{sec:discussion} we discuss the consequences of DBSCAN's behaviour on the synthetic set of clusters and make predictions for spectroscopic surveys. We summarize our conclusions in Section~\ref{sec:conclusions}.

\section{Synthetic clusters}
\label{sec:synthetic}

Our goal in this work is to test chemical tagging on a chemical space that closely emulates real data. We describe in this section the methods by which we generate a sample of stars, assigning each a synthetic spectrum, a set of abundance measurements, and a cluster assignment, in a way that mimics our expectations from real data. We specify the number of stars observed, the volume of space in which they are observed, the cluster mass function, and the level of measurement uncertainty in their spectra and derived chemistry. The choices for these parameters can change to reflect a variety of underlying physics and spectroscopic surveys. To perform our tests, we create cluster observations as they would appear to a survey like the Sloan Digital Sky Survey's (\citealt{Eisenstein2011}, \citealt{Blanton2017}) Apache Point Observatory Galactic Evolution Experiment (APOGEE-\citealt{Majewski2017}). To do this, we draw on observations from APOGEE's DR14 \citep{Abolfathi2018}. We discuss how our results extend to other surveys in \S\ref{sec:conclusions}.

\subsection{Survey-observed cluster members}
\label{sec:survey}

Our goal is to create a realistic sample of stars with a chemical space to investigate. We begin by assuming a survey volume. For the purposes of this work, we take the volume $V$ to be an annulus containing the Sun and centred on the Galactic Centre, with a width of $6$ kpc. To keep things simple, we assume that the stellar mass density of the Milky Way is constant at $0.05 M_{\odot}/\mathrm{pc}^3$ (see \citealt{Bovy2017} for a measurement of this quantity). For a given survey volume we convert this density into the total surveyed stellar mass $M_{\rm region}$. Assuming a Kroupa IMF \citep{Kroupa2001}, 
\begin{equation}
\xi(m) \propto
	\begin{cases} 
      	\left(\frac{m}{M_{\odot}}\right)^{-1.3} & 0.1 M_{\odot} \leq m\leq 0.5 M_{\odot} \\
      	\left(\frac{m}{M_{\odot}}\right)^{-2.3} & 0.5 M_{\odot}\leq m \leq 5 M_{\odot}
   	\end{cases}
\end{equation}
we determine the typical mass per star, $m_{\star}=0.6 M_{\odot}/\mathrm{star}$, which allows us to convert stellar mass into a corresponding number of stars. Thus the total number of stars in the survey volume is $N_{\rm region} = M_{\rm region}/m_{\star}$. We choose the number of stars observed in our hypothetical survey, $N_{\rm survey}$, and calculate the overall sampling rate of our observations, $\gamma = N_{\rm survey}/N_{\rm region}$ (approximately $2\times 10^{-6}$ for our fiducial sample of 50,000 stars). Since the actual rate at which cluster members are sampled for a given cluster will vary between clusters, we use the overall sampling rate to parameterize an exponential distribution. From this distribution, we draw a unique sampling rate for each individual cluster, so the $j$'th cluster has sampling rate $\gamma_j$. This creates a semi-realistic scenario in which some clusters are sampled more aggressively than others by random chance. The mass distribution of clusters is assumed to be given by a power law
\begin{equation}
	\xi(M) \propto \left(\frac{M}{M_\odot}\right)^{-\alpha}\,\, 50 M_{\odot} \leq M \leq 10^7 M_{\odot},
	\label{eqn:cmf}
\end{equation}
from \citet{Lada2003}, where cluster mass limits are taken from \citet{Ting2015}. We take $\alpha=2.1$ as our fiducial value of the power law index. The stellar mass for the $j$'th cluster observed by our simulated survey is given by the total cluster mass ($M_j$) multiplied by the sampling rate for that cluster ($\gamma_j$). We find the number of stars in each cluster $j$ as $N_j^{\star} = (\gamma_j M_j)/m_{\star}$.

By following this process, we choose for an imagined survey the volume $V$, the number of stars observed $N_{\rm survey}$, and the power law index of the cluster mass function $\alpha$, and obtain a simulated set of clusters. Each star produced has an integer label which indexes it to a birth cluster. 

\subsection{Creating clusters} 
\label{sec:creating}

\subsubsection{Cluster abundances}
\label{sec:centers}
We want to create cluster chemical signatures that have realistic abundance patterns, rather than drawing from simple distributions for each abundance. To do this, we consider the 15 abundances reported for each star in APOGEE DR12 \citep{Alam2015}: [C/H], [N/H], [O/H], [Na/H], [Mg/H], [Al/H], [Si/H], [S/H], [K/H], [Ca/H], [Ti/H], [V/H], [Mn/H], [Ni/H], and [Fe/H]. We make use of APOGEE's DR14, which reports abundance ratios with respect Fe rather than H, so we use [Fe/H] values for each star to obtain the abundance ratios with respect to H. Using the \texttt{apogee} package \citep{Bovy2016}, we collect APOGEE's observations as processed by the APOGEE Stellar Parameter and Chemical Abundances Pipeline (ASPCAP - \citealt{GarciaPerez2016}), which provide the 15 abundances. We make the conservative choice to cut any stars that has been given a non-zero \textsc{apogee\_starflag} bitmask value \citep{Holtzman2015}. We further remove any star that for any of the 15 elements listed above has a value of -9999, a value used by the ASPCAP to indicate no measurement was possible for that element with the observed spectrum. Finally, we cut any star missing a surface gravity ($\log g$) or effective temperature ($T_{\rm eff}$) measurement and restrict the sample to stars with $4700\,\mathrm{K} < T_{\rm eff} < 4900\,\mathrm{K}$ to capture only the red giant stars for which ASPCAP results are more accurate \citep{Holtzman2015}.

To create each cluster, we select a star at random from the APOGEE sample described above, and use its abundances as the centre of the cluster in abundance space. It is unlikely that any member star will have exactly the same abundances as the cluster centre (except in the case of zero uncertainty), but the centre will represent the typical abundances of the set of stars in that cluster.

\subsubsection{Synthetic abundances}
\label{sec:synabun}
Once cluster centres are set, we generate the properties of the member stars. We create a chemical signature for each member star in a cluster by choosing its abundances randomly from a 15-dimensional normal distribution centred on the abundances of the cluster centre. We use three different choices for the spreads in the normal distribution in each abundance dimension, motivated by the reported precision in abundances from APOGEE. We assume that the spread in each abundances is independent of the other abundances. The three cases we considered for the spread in abundances within a cluster are detailed in Table~\ref{tab:spreads} and summarized below:

\begin{itemize}
	\item \emph{conservative}: Member abundances selected with spread around cluster centre corresponding to uncertainties as quoted for APOGEE DR12 in \citet{Holtzman2015}.
	\item \emph{optimistic}: Member abundances selected with spread around cluster centre corresponding to uncertainties from the machine learning approach to abundance derivation from spectra at SNR=50 in \citet{Leung2019}. These spreads are similar to those from other data-driven approaches (e.g. \citealt{Casey2016a}).
	\item \emph{theoretical}: Member abundances selected with spread around cluster centre corresponding to theoretical uncertainties at the Cramer-Rao bound computed in \citet{Ting2016b}.
\end{itemize} 

\begin{table}
\centering
\caption{Uncertainties used to add normally distributed noise to cluster member abundances around the chosen centre. Conservative from \citet{Holtzman2015}, optimistic from \citet{Leung2019}, and theoretical from \citet{Ting2016b}.}
	\begin{tabular}{|r|l|l|l|}
	\label{tab:spreads}
		element & conservative & optimistic & theoretical \\
		\hline
		{[C/H]} & $3.5\times 10 ^{-2}$ & $3.3\times 10^{-2}$ & $5.0\times 10^{-3}$ \\
		{[N/H]} & $6.7\times 10^{-2}$ & $3.1\times 10^{-2}$ & $1.0\times 10^{-2}$\\
		{[O/H]} & $5.0\times 10^{-2}$ & $2.0\times 10^{-2}$ & $1.0\times 10^{-2}$ \\
		{[Na/H]} & $6.4\times 10^{-2}$ & $4.9\times 10^{-2}$ & $3.8\times 10^{-2}$ \\
		{[Mg/H]} & $5.3\times 10^{-2}$ & $1.7\times 10^{-2}$ & $7.6\times 10^{-3}$ \\
		{[Al/H]} & $6.7\times 10^{-2}$ & $3.3\times 10^{-2}$ & $2.0\times 10^{-2}$ \\
		{[Si/H]} & $7.7\times 10^{-2}$ & $1.8\times 10^{-2}$ & $8.2\times 10^{-3}$ \\
		{[S/H]} & $6.3\times 10^{-2}$ & $4.0\times 10^{-2}$ & $2.4\times 10^{-2}$ \\
		{[K/H]} & $6.5\times 10^{-2}$ & $2.1\times 10^{-2}$ & $4.4\times 10^{-2}$ \\
		{[Ca/H]} & $5.9\times 10^{-2}$ & $1.5\times 10^{-2}$ & $1.6\times 10^{-2}$ \\
		{[Ti/H]} & $7.2\times 10^{-2}$ & $1.9\times 10^{-2}$ & $1.8\times 10^{-2}$ \\
		{[V/H]} & $8.8\times 10^{-2}$ & $3.9\times 10^{-2}$ & $6.0\times 10^{-2}$ \\
		{[Mn/H]} & $6.1\times 10^{-2}$ & $2.2\times 10^{-2}$ & $1.3\times 10^{-2}$ \\
		{[Ni/H]} & $6.0\times 10^{-2}$ & $1.7\times 10^{-2}$ & $1.0\times 10^{-2}$ \\
		{[Fe/H]} & $5.3\times 10^{-2}$ & $1.5\times 10^{-2}$ & $4.3\times 10^{-4}$
	\end{tabular}	
\end{table}

Considered together, the three possible choices of abundances for each star constitute three abundance spaces on which we perform chemical tagging. However, we do not use the member abundances to generate a spectrum for each star, as the relationship between uncertainty in stellar abundance and uncertainty in flux measurement is somewhat obscure. Our process for generating member spectra is described in the following section.

\subsubsection{Synthetic spectra}
\label{sec:synspec}

Separate from the assignment of abundances, we create spectra for each of the member stars in a cluster, assuming that all members truly share the abundances of their cluster centre. In this case, spread in the abundance measurements within a cluster is a combination of uncertainty in the spectra and uncertainty in the method with which abundances are derived. Assuming that each member has the abundances of the cluster centre, we select a $T_{\rm eff}$ and $\log g$ for each member star by assigning it the properties of a random star from the APOGEE sample described in \S\ref{sec:creating} above. This method allows us to preserve the relationship between $T_{\rm eff}$ and $\log g$ observed in real stars. Stellar spectra are created using the polynomial spectral model developed in \citet{Rix2016}, as this allows us to quickly generate large stellar samples. The model produces spectra that emulate the spectra produced using spectral synthesis with model atmospheres. We add normally distributed noise to the spectra, at the order of one hundredth of the signal. This noise represents the observational uncertainties in each spectrum.

Since these synthetic spectra have non chemical photospheric information ($T_{\rm eff}$ and $\log g$), they cannot be used in raw form for chemical tagging.  To remove the influence of the photospheric properties on the spectra, we employ the same polynomial fitting technique used in \citet{Price-Jones2018}. In this approach, we correct for $T_{\rm eff}$ and $\log g$ at each wavelength independently. Once a wavelength is selected, we use the $T_{\rm eff}$ and $\log g$ assigned when we created our sample in a quadratic fit to the flux at that wavelength across all stars in the sample. We then subtract that fit from the flux and repeat with the next wavelength until the flux values at all wavelengths are corrected. The residuals of the fits constitute a corrected set of spectra where photospheric effects have been removed, ideally leaving only variations due to the differences in stellar chemistry and the noise we inserted. We refer to the fit residuals as the \emph{unmodified} spectra henceforth, since they receive no further processing.

\subsubsection{Projecting the spectra}
\label{sec:project}
\begin{figure}
\centering
\includegraphics[width = \linewidth]{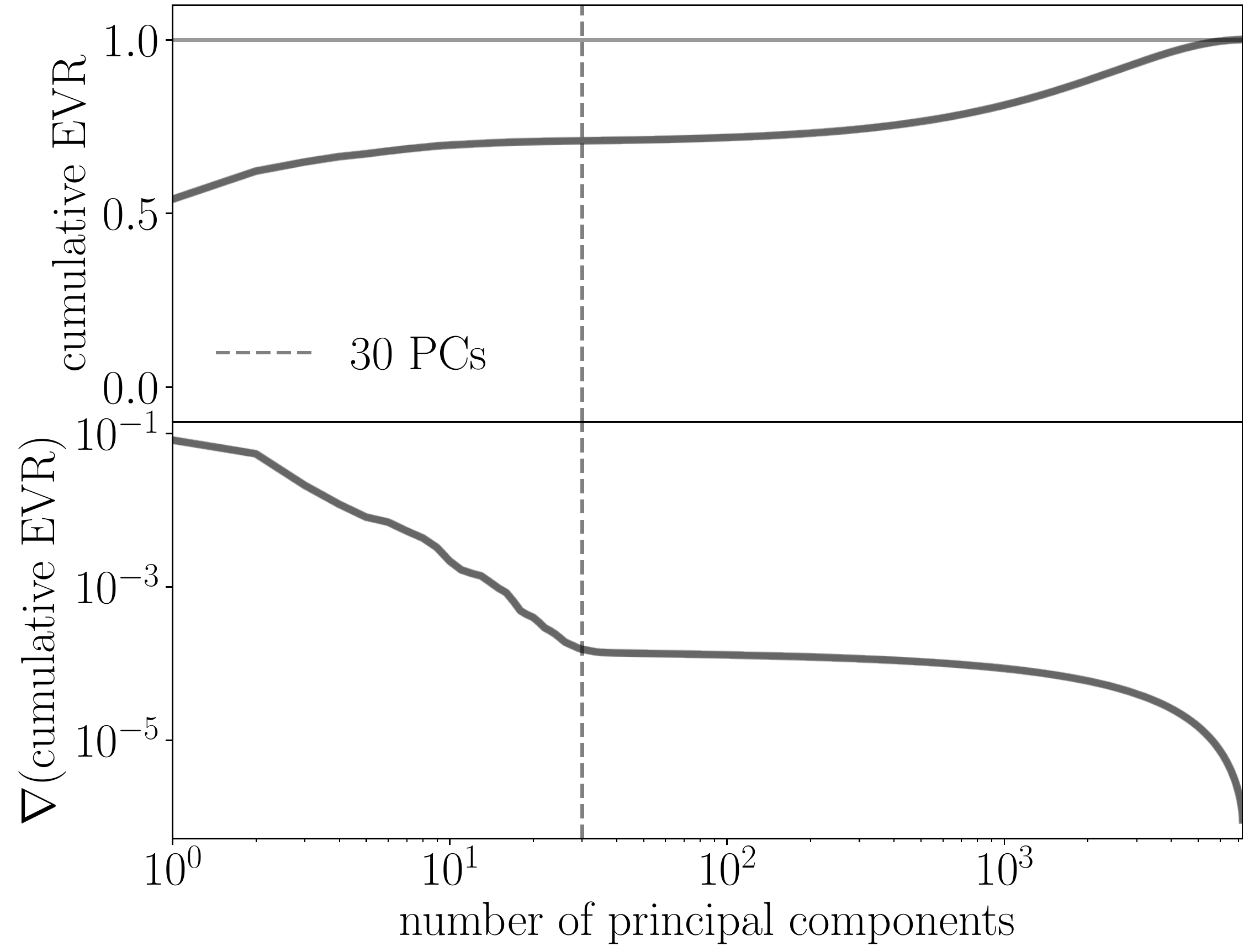}
\caption{\textit{Top panel:} Cumulative explained variance ratio as a function of the number of principal components used to model a sample of 30,000 spectra. \textit{Bottom panel:} Gradient of the cumulative explained variance ratio as a function of the number of principal components. Beyond 30 principal components, adding additional principal components explains noise rather than variance in the spectra, as can be see by the abrupt levelling off of the gradient at that point.}
\label{fig:PCs}
\end{figure}

The spectra are high-dimensional, spanning over seven thousand pixels, and many of these pixels are dominated by continuum flux and observational noise instead of chemical information. To emphasize the contribution of pixels with the greatest variation between spectra, we create a new chemical space by projecting the spectra into fewer dimensions before implementing a cluster finding algorithm. We do this using principal component analysis (PCA; \citealt{Joliffe2002}, \citealt{Ivezic2014}). PCA decomposes the spectra into a linear combination of principal components (PCs) by solving for the eigenvectors of the covariance matrix of the data.  

We choose to use 30 PCs to reduce the dimensionality of the spectra. Figure~\ref{fig:PCs} provides the rationale for this choice. A full PCA reduction will solve for PCs up to the minimum of the number of data dimensions and the number of observables - with that many PCs, the data are perfectly reproducible. We computed the full suite of PCs for a sample of 30,000 stars, as well as their associated explained variance ratio (EVR), given by
\begin{equation}
	\mathrm{EVR}(i) = \frac{\sigma^2_{\rm model}(i)}{\sigma^2_{\rm data}}
\end{equation}
where $\sigma^2_{\rm model}(i)$ is the variance in the model for the data constructed of the $i$'th PC and $\sigma^2_{\rm data}$ is the variance in the original data. The PCs are ordered by the amount of variance they account for, with low $i$ PCs accounting for most of the variance, and so this quantity decreases with increasing $i$.

 The cumulative distribution of the EVR and its gradient are shown in the top and bottom panels of Figure~\ref{fig:PCs}, respectively. At the maximum possible number of PCs, the cumulative EVR goes to 1, indicating that, as expected, all variance can be explained with this many PCs. However, it is not necessary to use the full possible suite of PCs to get a sufficiently accurate model of the data, especially since we construct the data to include both noise that we have no desire to model and continuum pixels with no chemical information. In Figure~\ref{fig:PCs}, there is a plateau in both the cumulative EVR and its gradient at 30 PCs. Beyond this point, significant improvements in the variance explained require an order of magnitude increase in the number of PCs. Additional PCs beyond the first 30 are responsible for adding variations to the model that will only emulate noise or continuum variations in our synthetic spectra. Therefore, we choose to use 30 PCs to reduce the spectra into chemistry-relevant directions, reducing the influence of the noise and continuum emission pixels in this new chemical space.

We refer to the case where spectra are projected along their first 30 PCs as \emph{principal components}.  This projection is a data driven approach to the challenge of reducing the importance of noise when clustering stars according to their spectra.

\section{Identifying clusters}
\label{sec:identify}

The goal of this study is to identify a clustering algorithm that can find groups of stars in high dimensional chemical space without needing information about how many groups are expected. This makes Density Based Spatial Clustering of Applications with Noise (DBSCAN - \citealt{Ester1996}) a logical choice of clustering algorithm, as it relies only on the density of stars in chemical space to locate groups, without requiring any prior information on the number of groups present.

In general, we describe a star $s$ as having a position in chemical space $\textbf{x}_s$, where $\textbf{x}_s$ may be an list of abundances (for an abundance-based chemical space) or a list of fluxes (for spectrum-based chemical spaces). Stars are indexed from 1 to $N_{\rm survey}$, and the function used to compute pairwise distances between any two stars is denoted with $\mathcal{D}$. While any distance metric can be chosen, we use a Euclidean metric. We summarize these and other relevant symbols in Table~\ref{tab:params} in the order in which they first appear in this paper.

To differentiate them from our synthetic clusters, we henceforth refer to the stars classified as belonging together by DBSCAN as `groups', and save the designation `clusters' to refer to the known membership of the stars. 

\subsection{DBSCAN}
\label{sec:dbscan}
DBSCAN classifies stars as `core stars', `border stars', and `noise stars', where the former two types are members of groups but the latter is a label for outlying stars. Each star's type depends on the input choice of $\epsilon$ and $N_{\rm pts}$, the quantities that parameterize DBSCAN. The region of chemical space that is within distance $\epsilon$ of a star is that star's `$\epsilon$-neighbourhood'. If a star has $N_{\rm pts}$ stars within its $\epsilon$-neighbourhood, it is considered a core star. Any point that is within a core star's $\epsilon$-neighbourhood but is not itself a core star, is labelled a border star. Any star that is not in the $\epsilon$-neighbourhood of any core star is a noise star.

The DBSCAN algorithm begins by assuming all stars are noise stars. The algorithm steps through each star, and for a given choice of $\epsilon$ and $N_{\rm pts}$, checks whether it is a core star. Once the core stars are identified, the algorithm organizes the stars into groups, by considering each star sequentially. If the star is a core star, and is in the $\epsilon$-neighbourhood of another core star with a group label, that star is given the same group label. If it is not, the core star is given a new group label. If the star is not a core star, the algorithm determines whether it is a border star (in which case it is assigned to the group its neighbouring core point belongs to) or a noise star. The method is summarized below as algorithm~\ref{alg:DBSCAN}.

\begin{algorithm}
\caption{Summary of DBSCAN as applied to stellar chemical space.}
\label{alg:DBSCAN}
\SetAlgoLined
\KwData{A chemical space, either abundances or processed spectra for each star.}
\KwResult{Integer labels for each star, assigning them to a group if they are $\geq 0$ or labelling them as noise if the integer is -1}
compute the pairwise distances between all stars\;
choose $\epsilon$ and $N_{\rm pts}$ values\;
initialize a list containing the group label for each star, and set them all to -1, the flag for noise stars\;
\For{each star $s$}{
find the number of stars in its $\epsilon$-neighbourhood, $N_{\epsilon}^s$\;
\uIf{$N_{\epsilon}^s \geq N_{\rm pts}$}{
label $s$ as a core star\;
}
\uElseIf{$N_{\epsilon}^s < N_{\rm pts}$}{
leave $s$ as a noise star\;
}
}
\For{each star $s$}{
\uIf{$s$ is a core star}{
\uIf{there are no core stars within the $\epsilon$-neighbourhood of $s$}{
give $s$ the next unused integer for a group label ($s$ is a \emph{core star} in a new group)\;
}
\uElseIf{there is another core star $s'$ within the $\epsilon$-neighbourhood of $s$}{
give $s$ the same group label as $s'$ ($s$ is a \emph{core star} in an existing group)\;
}
}
\uElseIf{$s$ is a noise star}{
\uIf{$s$ is in the $\epsilon$-neighbourhood of a core star $s'$}{
give $s$ the same group label as $s'$ ($s$ is a \emph{border star} in an existing group)\;
}
\uElse{
leave $s$ with the \emph{noise star} label\;
}
}
}
return the group label for each point\;

\end{algorithm}

Note that because DBSCAN considers the points in the order they are given, changing the order of the dataset can lead to different group assignments. For core points, only the label of the cluster will change, while membership remains consistent. However border points may be assigned to different clusters depending on data order.

In this work, we use the DBSCAN algorithm implemented in the \texttt{scikit-learn} Python package \citep{scikit-learn}. We also define $\epsilon$ in terms of what we call the normalized $\epsilon$, $\tilde{\epsilon}$:
\begin{equation}
	\epsilon = \tilde{\epsilon} \cdot \mathrm{median}\left(\overset{N_{\rm survey}}{\underset{s,t}{\mathcal{D}}}(\mathbf{x}_s,\mathbf{x}_t)\right).
	\label{eqn:eps}
\end{equation}
where $\tilde{\epsilon}$ is a factor between 0 and 1 that scales the median of all pairwise distances between stars in chemical space.

\subsection{External cluster validation}
\label{sec:external}
When clusters are known, we assess DBSCAN's performance with homogeneity and completeness scores for each group $g_i$ it finds when compared to each original cluster $o_j$. Homogeneity for group $g_i$ is defined as
\begin{equation}
	H_i^j = \frac{\mathrm{number\,of\,stars\,in\,}g_i\mathrm{\,from\,}o_j}{\mathrm{number\,of\,stars\, in\,}g_i}
\label{eqn:H}
\end{equation}
If group $g_i$ contains only members of a single original cluster $o_j$, then $H_i^j$ gives a perfect homogeneity score of 1. 

This score alone is not enough; if DBSCAN splits large clusters into small groups, each might have a corresponding $o_j$ that gives a perfect homogeneity score, but the groups would poorly represent the input clusters. In addition to homogeneity scores, we also compute completeness scores for each group, defined as
\begin{equation}
	C_i^j = \frac{\mathrm{number\,of\,stars\,in\,}g_i\mathrm{\,from\,}o_j}{\mathrm{number\,of\,stars \,in\,}o_j}
\label{eqn:C}
\end{equation}
If all members of $o_j$ are assigned to the same $g_i$, then $C_i^j$ gives a perfect completeness score of 1. 

For each group $g_i$, we assign a single homogeneity $H_i$ and completeness $C_i$ score by choosing the compared original cluster to be the one that contributes the most stars to $g_i$. We call this original cluster the `matched cluster'. This method permits the same original cluster to be matched to multiple groups, and has the effect of ensuring the homogeneity score for each group takes on a maximal value. However, as the number of groups matched with the same original cluster increases, the completeness score for each group will decrease.

\subsubsection{Recovery fraction}
\label{sec:recovery}

Unlike the metrics discussed above, the recovery fraction for a given iteration of DBSCAN is not computed on a per cluster basis, but is a way to assess the overall performance of the algorithm across all input data for a given parameterization. While homogeneity and completeness provide good overall estimates of the ability of DBSCAN to reproduce the initial clustering, we need a combined score to ensure that individual clusters are identified properly. To compute this combined score, we first randomly choose a set of the original clusters that meet a size threshold $O = \{o_j\}$. We typically choose 10 clusters that meet a size threshold of 15 members. These numbers approximate the number and membership of open clusters we might use to validate clustering in observed data (e.g. \citealt{Frinchaboy2003}).  We then find the set of DBSCAN identified groups that these clusters were matched to, $\{g_i\}$.


We are only interested in the fraction of clusters recovered successfully. To assess the quality of recovery, we place lower limits on the homogeneity and completeness scores for the groups. A cluster is considered to be `recovered' if the group it is matched to exceeds our homogeneity and completeness thresholds, and so we form a subset of $\{g_i\}$ called $R$ that consists only of `recovered' clusters. Therefore our recovery fraction is
\begin{equation}
	\mathrm{RF} = \frac{\mathrm{number\,of\,groups\,in\,}R}{\mathrm{number\,of\,clusters\,in\,}O},
\label{eqn:RF}
\end{equation}
and in this work we typically set our homogeneity and completeness thresholds to be 0.7. To get robust values for $\mathrm{RF}$, we make many random selections for $O$, find the matching $R$, and compute the corresponding $\mathrm{RF}$, taking the median of these as our overall recovery fraction.

Increasing the homogeneity and completeness thresholds results in only considering groups with very high fidelity to the original cluster as true recovery, and decreases RF. By making appropriate threshold sizes we can test how well clusters of a given size might be found in observed data.

\begin{figure}
\centering
\includegraphics[width = \linewidth]{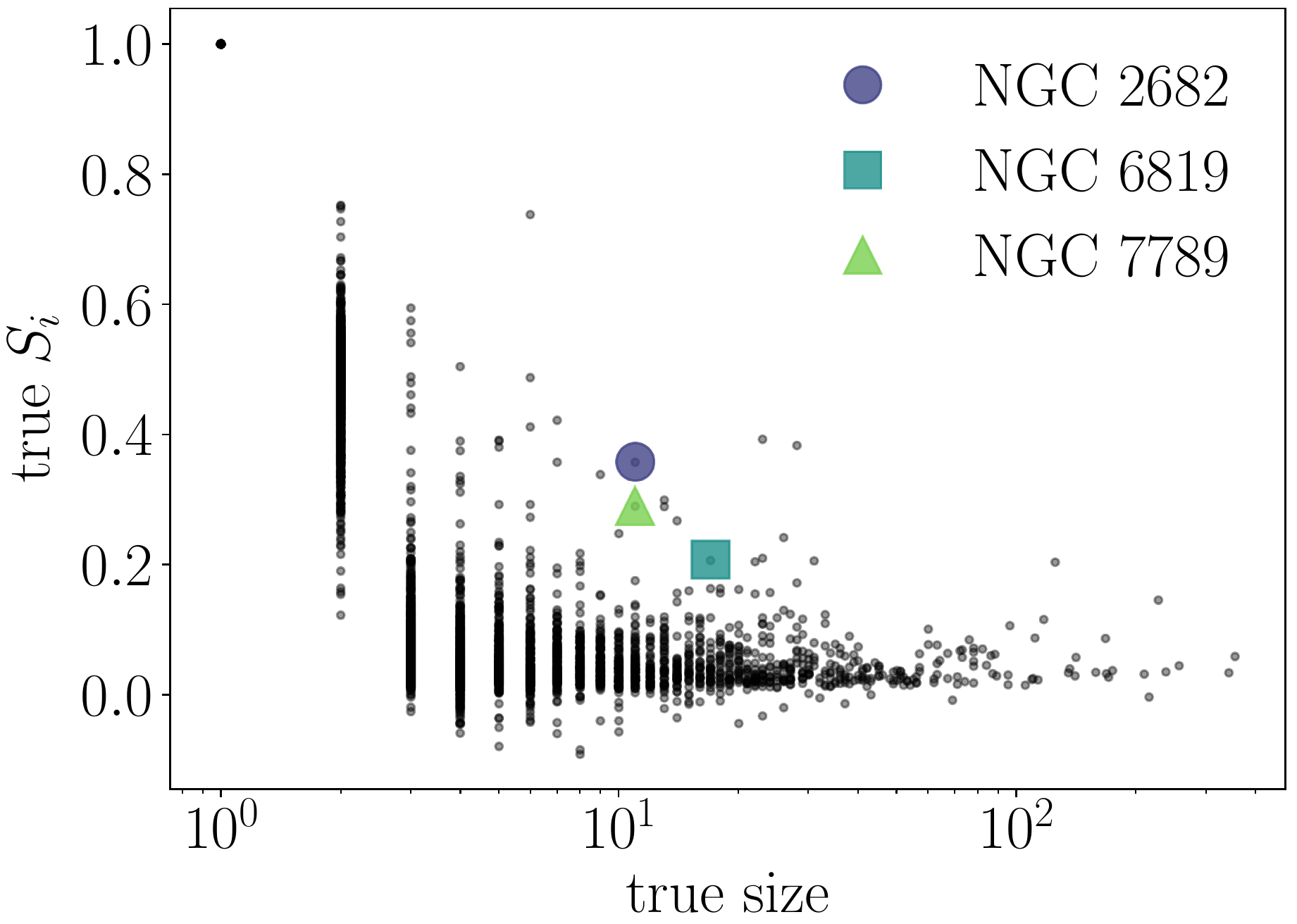}
\caption{The distribution of silhouette coefficients $S_i$ (equation~\eqref{eqn:silhouette}) for our input clusters in principal component space as a function of the number of stars in the input clusters. Three open clusters identified in by OCCAM \citep{Donor2018} are marked with coloured symbols for comparison. At low sizes, $S_i$ can take on higher values, since it is easier for a small cluster with normally distributed members to be compact relative to its background (note that all single-membered clusters have $S_i$ of exactly one). The open clusters are typically more distinct from their surroundings than our simulated clusters of their size.}
\label{fig:tsils}
\end{figure}

\subsection{Internal cluster validation}
\label{sec:internal}

In addition to these external methods, we need a validation criterion that can evaluate the performance of DBSCAN even when true clusters are unknown, because this is the situation we face when using real data. Criteria of this kind often rely on comparing the distances between groups members to the distances between groups. One well known choice for this is the Silhouette Coefficient \citep{Rousseeuw1987}. This metric compares the pairwise distances between group members (internal distance) to the pairwise distances between group members and nearby non-group members (external distance).  

The stars in our sample have positions $\{\mathbf{x}_s\}$ in spectral or chemical space. We compute the internal distance as the typical distance between members of the same group. For the $i$'th group $g_i$, this internal distance is
\begin{equation}
	d_{\rm int}^i = \mathrm{median}\left(\underset{s,t\in g_i,s\neq t}{\mathcal{D}}\left(\textbf{x}_s,\textbf{x}_t\right)\right)
\end{equation}
where $\mathcal{D}$ is the operator that calculates the distance between each pair of points and $s$ and $t$ index the members of $g_i$. 

To calculate the overall external distance for group $g_i$, we find the set of distances between the members of $g_i$ and members of nearby group $g_j$, repeating for other nearby $g_j$. For our computation, of $d_{\mathrm{ext}}^i$, we use as the $g_j$'s the twenty groups with median positions closest to the median position of $g_i$. The median of these distances is the external distance.
\begin{equation}
	d_{\rm ext}^{i} = \mathrm{median}\left(\underset{s\in g_i, t\in g_j}{\mathcal{D}}\left(\textbf{x}_s,\textbf{x}_t\right)\right)_{\mathrm{nearby}\,g_j}
\end{equation}
We define the silhouette coefficient for the $i$'th group as
\begin{equation}
	S_i = \frac{d^{i}_{\rm ext} - d^{i}_{\rm int}}{\max\left(d^{i}_{\rm ext},d^{i}_{\rm int}\right)}.
\label{eqn:silhouette}
\end{equation}
For a very distinct group, $d_{\rm int}^{i}$ will be much less than $d_{\rm ext}^{i}$, and the silhouette coefficient reaches a maximum value of 1. We show the typical distribution of silhouette coefficients for our input data in Figure~\ref{fig:tsils}. As group size increases, so too does the chance that groups will overlap, leading to a decrease in silhouette coefficient with group size. In the following section, we investigate the success with which DBSCAN can identify these groups.

Note that although we have defined the silhouette coefficient for a group, it can just as easily be computed for clusters in the initial dataset.

In Table~\ref{tab:params} we summarize the symbols and quantities defined in the preceding sections. In the following section, we describe the results of applying DBSCAN to our synthetic data, using our metrics to assess the success of the algorithm.

\begin{table}
\centering
\caption{Symbols used in our methods and their meaning, presented in the order in which they first appear in this paper.}
	\begin{tabular}{|r|l|c|}
	\label{tab:params}
	Symbol & Definition & Section \\
	& & \& Equation\\
	\hline
	$V$ & volume of physical space surveyed & \S\ref{sec:survey} \\
	$N_{\rm survey}$ & number of stars in the chemical space & \S\ref{sec:survey} \\
	$\alpha$ & index of the cluster mass function & \S\ref{sec:survey}-\eqref{eqn:cmf} \\
	\rule{0pt}{4ex}  $\mathbf{x}_s$ & position of star $s$ in chemical space & \S\ref{sec:identify} \\
	$\mathcal{D}$ & function that computes pairwise distances & \S\ref{sec:identify} \\
	& between stars in chemical space & \\
	$\epsilon$ & radius of a star's `$\epsilon$-neighbourhood' in & \S\ref{sec:dbscan} \\
	& chemical space& \\
	$N_{\rm pts}$ & minimum number of stars needed in a & \S\ref{sec:dbscan}\\ 
	& star's $\epsilon$-neighbourhood for it to be &\\
	& considered a `core star' & \\
	$\tilde{\epsilon}$ & the normalized $\epsilon$ factor that & \S\ref{sec:dbscan}-\eqref{eqn:eps} \\
	& converts the median pairwise distance &\\
	& between stars into $\epsilon$& \\
	\rule{0pt}{4ex} $g_i$ & the $i$'th group identified by DBSCAN & \S\ref{sec:external} \\
	$o_j$ & the $j$'th original cluster in the initial dataset & \S\ref{sec:external} \\
	\rule{0pt}{4ex} $H_i$ & homogeneity of $g_i$ when compared to the & \S\ref{sec:external}-\eqref{eqn:H}\\
	& original cluster that contributed the majority & \\
	& of its members &\\
	$C_i$ & completeness of $g_i$ when compared to the & \S\ref{sec:external}-\eqref{eqn:C}\\
	& original cluster that contributed the majority & \\
	& of its members &\\
	RF & fraction of clusters recovered with threshold  & \S\ref{sec:recovery}-\eqref{eqn:RF} \\
	& for minimum homogeneity and completeness & \\
	$S_i$ & silhouette coefficient for group or cluster $i$ & \S\ref{sec:internal}-\eqref{eqn:silhouette} \\
	
	\end{tabular}
\end{table}

\section{Results}
\label{sec:results}

Throughout this section, we explore a variety of chemical spaces (described in \S\ref{sec:creating}), which we list below for reference.

\begin{itemize}
	\item \emph{spectra}:
	\begin{itemize}
		\item \emph{unmodified} - Synthetic spectra after the subtraction of a fit to each pixel in $T_{\rm eff}$, $\log g$ and [Fe/H].
		\item \emph{principal components} - Spectra projected onto the first 30 principal components derived from PCA.
	\end{itemize}
	\item \emph{abundances} (from Table~\ref{tab:spreads}):
	\begin{itemize}
		\item \emph{conservative} - from \citet{Holtzman2015}
		\item \emph{optimistic} - from \citet{Leung2019}
		\item \emph{theoretical} - from \citet{Ting2016b}
	\end{itemize}
\end{itemize}

We begin by explaining our parameter choices to create an APOGEE-like survey, followed by a description of our parameter choice for DBSCAN. We summarize the results of our assessment statistics before describing the recovery fraction in more detail. 

\subsection{Creating an APOGEE-like simulation}

We have outlined our general approach to creating synthetic data in \S\ref{sec:creating}, and we describe here the specific choices we make to generate APOGEE-like synthetic data. The APOGEE survey releases all data publicly, and so we mimic the survey's derived abundances and stellar spectra. Note however that the intent of our simulation is not to exactly reproduce the APOGEE survey, but to use the same abundances and spectral window.

Our simplified approach does not account for the APOGEE selection function (see \S\ref{sec:assumptions} for more discussion of this), and we assume that stars observed by the survey are drawn from birth clusters that form in an annulus that contains the Sun and is centred on the Galactic Centre. This choice of volume allows our study to determine the capabilities of chemical tagging in the Milky Way disk. The annulus from which stars are drawn has a total height of 1 kpc, and we vary its width to observe the effects on clustering success, but use a 6 kpc width as a fiducial choice. For each choice of annulus width, we observe a sample of $5\times 10^4$ stars, or roughly $3\times 10^4$ $M_\odot$. Our simplifying assumptions on Milky Way density imply that there is $1.5\times 10^{10}$ $M_\odot$ of stellar mass in our fiducial survey volume, so in that case our sampling rate is $2\times 10^{-6}$.

Using this sampling rate and assuming the CMF has a power law index of $\alpha=2.1$ (see equation~\eqref{eqn:cmf}), we apply the process described in \S\ref{sec:survey}. We determine the number of members observed from each cluster in the volume until we have enough clusters to reach our target number of stars. For our fiducial sample of $5\times 10^4$ stars, this results in roughly $1.5\times 10^4$ clusters.

\begin{figure}
\centering
\includegraphics[width = \linewidth]{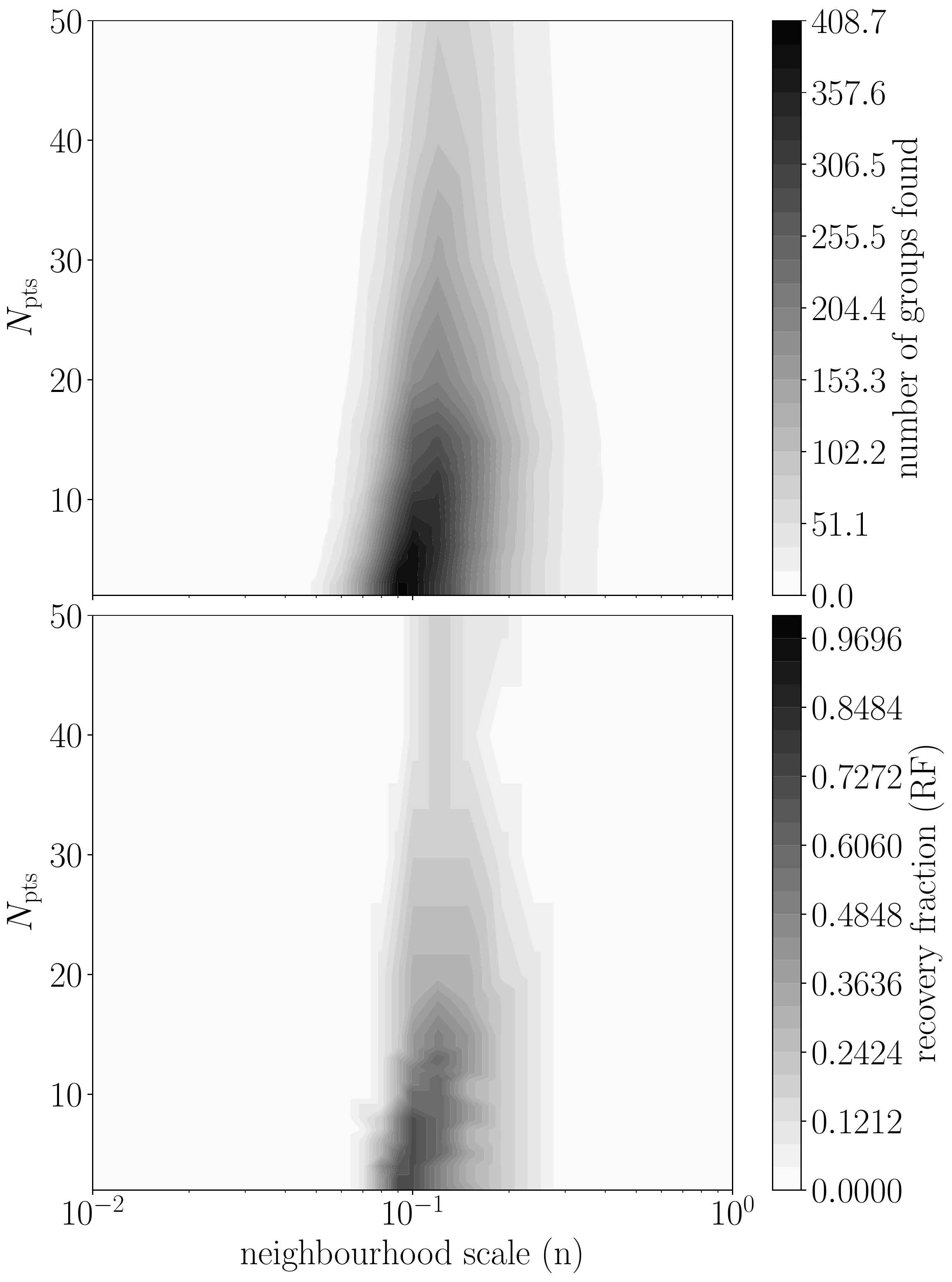}
\caption{Example contour plots showing the number of groups found (top panel) and the recovery fraction (bottom panel - see \S~\ref{sec:recovery}), as a function of the normalized $\epsilon$ ($\tilde{\epsilon}$) and the number of points in a neighbourhood ($N_{\rm pts}$) for our fiducial mock data sample. Note that by definition, $N_{\rm pts}$ is only allowed integer values, so values between points are interpolated with Delaunay triangulation \citep{Okabe1992}. While a narrow region in $\tilde{\epsilon}$ is preferred to maximize recovery fraction, a range of $N_{\rm pts}$ are permitted. However, a lower $N_{\rm pts}$ maximizes the recovery fraction. This example is the result of using DBSCAN on \emph{theoretical} abundance space, which has a low level of chemical space spread within clusters.}
\label{fig:ex}
\end{figure}

\subsection{Choosing DBSCAN parameters}
\label{sec:dbparam}

As described in \S\ref{sec:dbscan}, DBSCAN relies on two parameters to define a high density region: the radius of the region $\epsilon$, and the required number of points within the region $N_{\rm pts}$. Different choices of these parameters result in different groupings of stars. We choose $\epsilon$ as the product of the median pairwise distance between stars and the normalized $\epsilon$ ($\tilde{\epsilon}$  - equation~\eqref{eqn:eps}). We vary $\tilde{\epsilon}$ and $N_{\rm pts}$ and examine typical clustering success. For all of the data types listed at the beginning of this section we find that evaluation metrics change in continuous ways with both parameters. 

An example of how the number of clusters found by DBSCAN and the recovery fraction vary with $\tilde{\epsilon}$ (and thus vary with $\epsilon$), as well as with $N_{\rm pts}$ is shown in Figure~\ref{fig:ex}. There is obviously a preferred value for $\tilde{\epsilon}$ that maximizes RF, typically between 0.1-0.2. In addition, the choice of $\tilde{\epsilon}$ that maximizes RF is valid for a range of choices in $N_{\rm pts}$. The drop off in both the number of groups found and the recovery fraction with increasing $N_{\rm pts}$ begins where $N_{\rm pts}$ is approximately the 95th percentile of cluster sizes. This is quite sensible: when the vast majority of input clusters are smaller than the chosen $N_{\rm pts}$, DBSCAN will be unable to identify them and this will consequently lower the fraction of clusters that can be recovered.

Figure~\ref{fig:ex} shows results for our \emph{theoretical} abundance space, but we see similar results for all the chemical spaces we consider. In particular, we find that decreasing $N_{\rm pts}$ increases the total number of groups found, which is expected given the nature of DBSCAN. However, it is also at these lower choices for $N_{\rm pts}$ where recovery fraction is maximized. In light of this, we keep our range in $\tilde{\epsilon}$ for subsequent runs, but fix $N_{\rm pts}$ at 3.

\begin{figure}
\centering
\includegraphics[width = \linewidth]{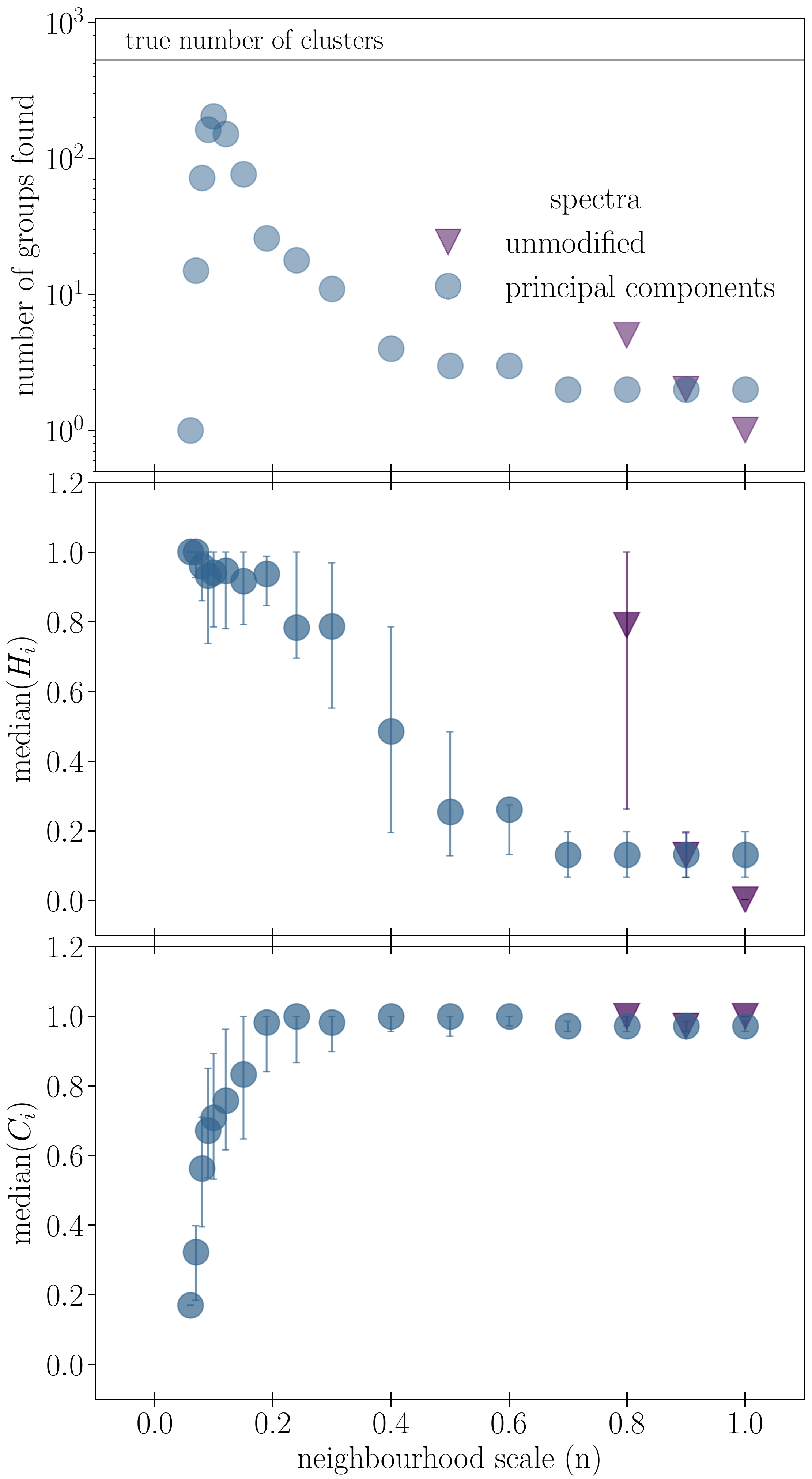}
\caption{\textit{Top panel}: The number of groups with more than 15 members found for the different spectra-based chemical spaces when using different values of $\tilde{\epsilon}$ in DBSCAN, where $\tilde{\epsilon}$ is the DBSCAN parameter that defines the neighbourhood of each star. \textit{Middle panel}: The median homogeneity (equation~\eqref{eqn:H}) of groups in the top panel. \textit{Bottom panel}: The median completeness (equation~\eqref{eqn:C}) of groups in the top panel. There is a particular value of $\tilde{\epsilon}$ that maximizes the number of clusters recovered, which corresponds to high median homogeneity and completeness. }
\label{fig:statspec}
\end{figure}

\begin{figure}
\centering
\includegraphics[width = \linewidth]{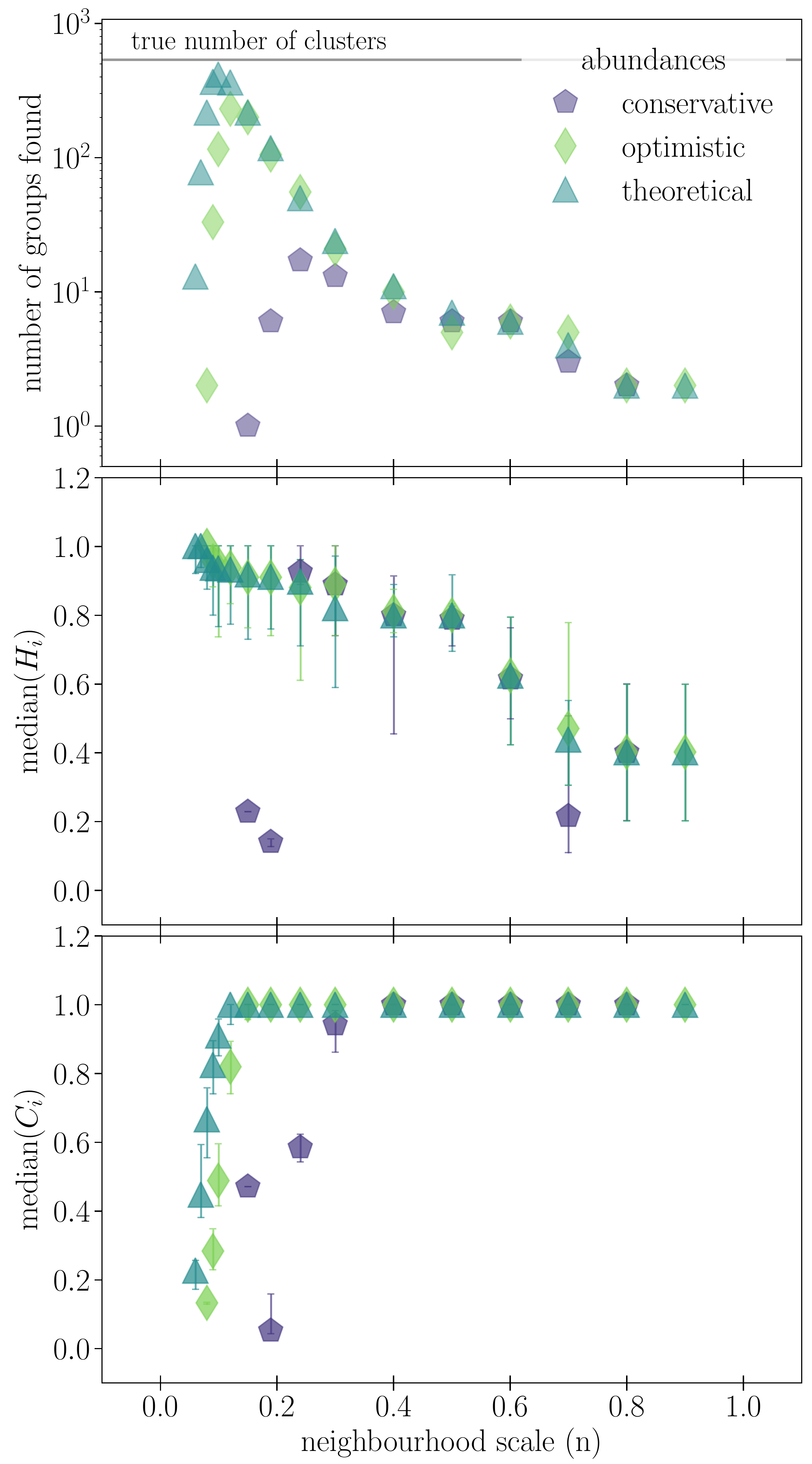}
\caption{Like Figure~\ref{fig:statspec}, but for chemical spaces based in abundances. As in Figure~\ref{fig:statspec}, there is a preferred choice of $\tilde{\epsilon}$ that maximizes the number of clusters recovered, and this $\tilde{\epsilon}$ corresponds to high median homogeneity and completeness. Also as in Figure~\ref{fig:statspec}, the median homogeneity and completeness are roughly inversely proportional.}
\label{fig:statabun}
\end{figure}

\begin{figure*}
\centering
\includegraphics[width = \linewidth]{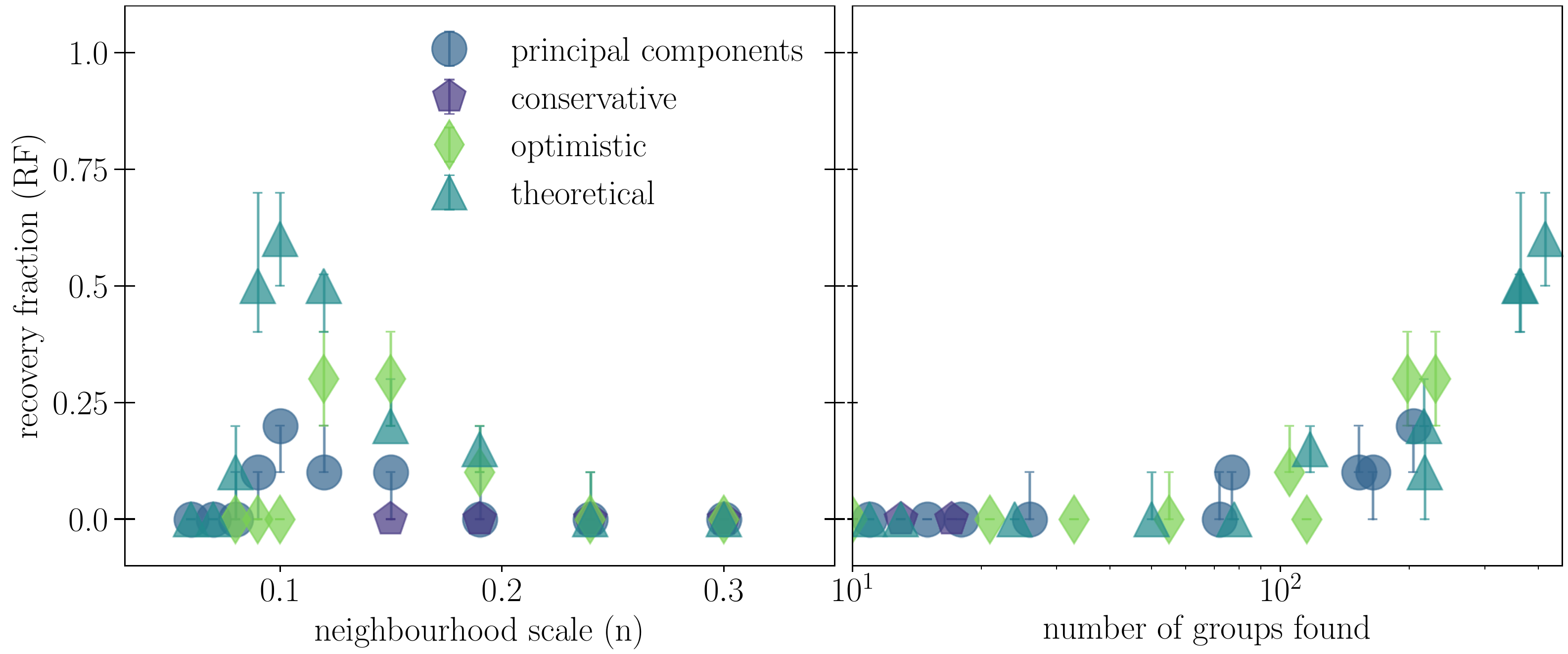}
\caption{Recovery fraction (equation~\eqref{eqn:RF}) as a function of normalized $\epsilon$ ($\tilde{\epsilon}$ - left panel) and the number of groups identified by DBSCAN (right panel). We show only results for chemical spaces that achieved non-zero recovery fraction: the principal component projection of the spectra and all three of the abundance chemical spaces. As with the number of clusters recovered in Figures~\ref{fig:statspec} and~\ref{fig:statabun}, there is a preferred value of $\tilde{\epsilon}$ that maximizes the recovery fraction. The right panel demonstrates the clear relationship between the number of clusters recovered and the recovery fraction, and so applications to real data need only choose the $\tilde{\epsilon}$ that returns the most clusters.}
\label{fig:recv}
\end{figure*}

\subsection{Summary statistics}
\label{sec:summary}

In \S\ref{sec:external}, we defined homogeneity and completeness as ways to validate the groups found by DBSCAN. To summarize the performance of DBSCAN across several parameter choices, we compute the quartiles of those quantities across all groups with more than 15 members identified by DBSCAN. Figures~\ref{fig:statspec} (the chemical spaces created with spectra) and~\ref{fig:statabun} (the chemical spaces created with abundances) show the results as a function of normalized $\epsilon$. Distances in all chemical spaces are computed with a Euclidean metric, and we do not apply normalization to any dimension. All dimensions in any given chemical space are measured in the same units, so variation along each dimension is of roughly the same order. However, our decision not to normalize the span of each dimension preserves the fact that some elements may be more informative than others when distinguishing between clusters. 

The top panel in each figure shows the total number of groups with more than 15 members found by DBSCAN for a given choice of $\tilde{\epsilon}$ as defined in equation~\eqref{eqn:eps}. The middle and bottom panels show the median homogeneity and completeness, respectively, while errorbars mark the positions of first and third quartile. Points for a particular data type are missing when that choice of $\tilde{\epsilon}$ did not return more than five groups.

We see that there is a preferred choice of $\tilde{\epsilon}$ for each type of data considered, with a clear relation between $\tilde{\epsilon}$ and the number of clusters recovered. Each data type also demonstrates the inverse relation between the median homogeneity and completeness scores; as $\tilde{\epsilon}$ increases, homogeneity decreases and completeness increases. This is because expanding the region considered when determining core points by increasing $\tilde{\epsilon}$ makes it more likely that a lower density cluster will be correctly identified (improving completeness), while simultaneously allowing more non-member interlopers (reducing homogeneity).

\subsection{Maximizing recovery fraction}

We compute a recovery fraction for each choice of $\tilde{\epsilon}$, shown in Figure~\ref{fig:recv}. We randomly select 10 clusters with more than 15 members from the input data, as this is the order of the number of known open clusters we might expect in a real dataset. We compute the recovery fraction (equation~\eqref{eqn:RF}), imposing homogeneity and completeness thresholds of 0.7. We repeat the random selection 100 times and use the first and third quartile in the resulting distribution of recovery fractions to produce an estimate of the scatter around the median resulting from using only 10 clusters. Comparing the the left panel with Figures~\ref{fig:statspec} and~\ref{fig:statabun} shows that while median homogeneity and completeness scores may be high for a given value of $\tilde{\epsilon}$, this does not necessarily imply a high recovery fraction. However, the peak in the recovery fraction with $\tilde{\epsilon}$ does correspond to the peak in the total number of clusters recovered, as shown in the right panel of Figure~\ref{fig:recv}.

\begin{figure*}
\centering
\includegraphics[width = \linewidth]{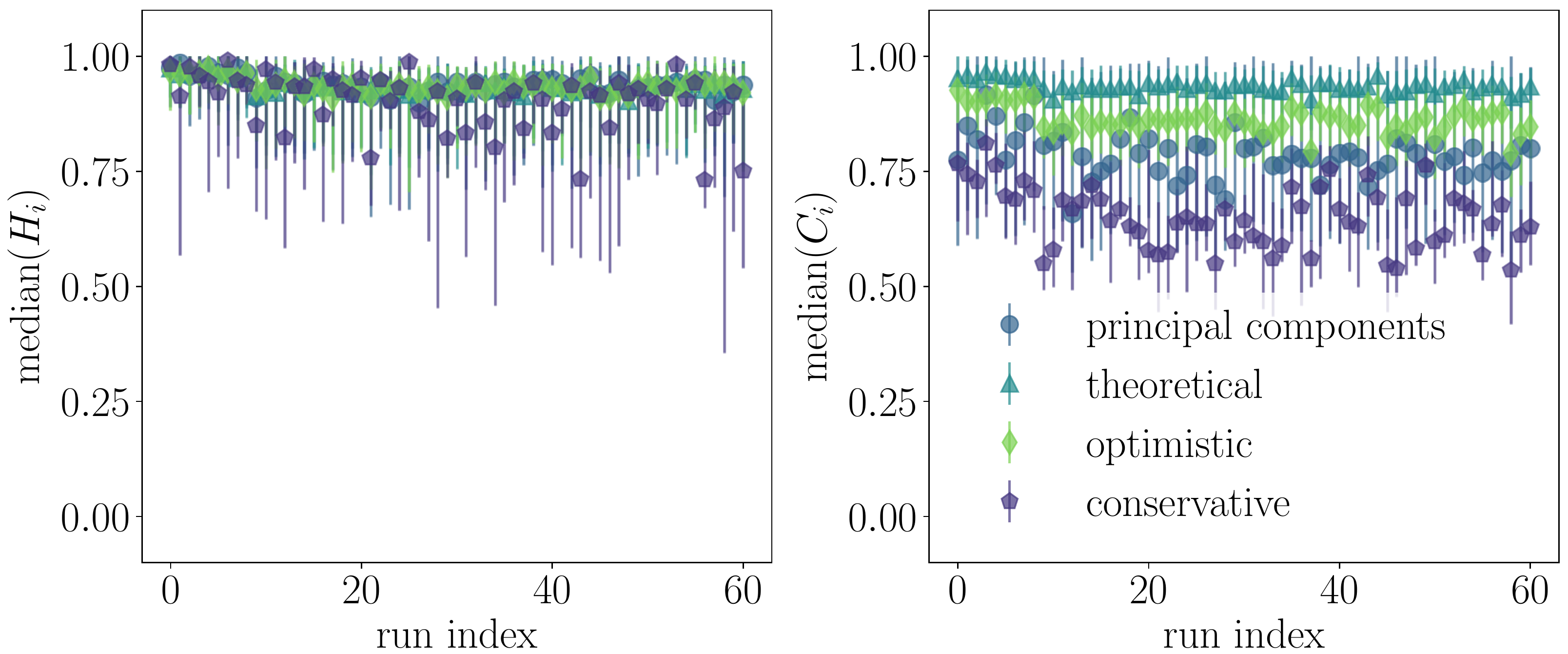}
\caption{Median homogeneity and completeness of groups found by DBSCAN with fixed $N_{\rm pts} = 3$ and the best $\tilde{\epsilon}$ choice (maximized recovery fraction) for each type of data listed at the top of \S\ref{sec:results}. Errorbars mark the locations of the first and third quartile. The same parameters were used for 61 runs of DBSCAN for different mock stellar samples, and we find consistent results for these statistics over all runs.}
\label{fig:averageprop}
\end{figure*}

To confirm that our statistics are not sensitive to the specifics of a particular simulation, we create dozens of simulated cluster samples with the same properties and run DBSCAN on each data type with the parameter choice that maximizes the recovery fraction for that type. The median homogeneity and completeness of the resulting groups are shown in Figure~\ref{fig:averageprop} and are relatively consistent across the 61 runs used. The different data types are split in the right hand plot of median completeness, clearly demonstrating that using DBSCAN on the abundance space with the largest intra-cluster spread has, as expected, the poorest performance. However, choosing the parameter set that maximizes recovery fraction for a given data type consistently provides excellent homogeneity and completeness scores.

\subsubsection{Recovery fraction for observed clusters}
\label{sec:ocrecv}
	
In observations, it will not be possible to compute a recovery fraction given that true cluster membership is unknown. However, it may be possible to compute a representative recovery fraction using known open clusters. Open clusters have the chemical properties we assume for our synthetic clusters, and so their recovery fraction should be related to the recovery fraction in the sample as a whole. However, creating a sample of known open clusters with good abundances is challenging. The Open Cluster Chemical Abundances and Mapping survey (OCCAM - \citealt{Frinchaboy2003}), has recently released updated membership lists in \citet{Donor2018} for APOGEE stars in open clusters. We apply quality cuts to ensure that all stars under consideration have good quality measurements for the 15 chemical abundances we generate for our simulated stars (\S\ref{sec:creating}). 

We also make use of the APOGEE spectra for the open cluster members. Observed spectra suffer from contamination of the underlying signal due to factors like atmospheric absorption lines, which are flagged in the APOGEE bitmask \citep{Holtzman2015}. Preparing these open cluster spectra for our algorithm is a useful way to prototype how our algorithm will work on larger observed samples. DBSCAN requires continuous data, with measurements in every dimension (pixel) for each star, but bitmask flagging indicates that some APOGEE pixels cannot be trusted. To apply DBSCAN to real spectra it is necessary to precompute the distances between spectra, for which there are three reasonable options. (1) Compute distances between spectra by comparing only pixels that were unmasked in both of the pair of spectra, then normalize those distances by the total number of pixels considered. (2) Compute distances on a subspace by limiting the number of pixels to ones where all spectra had good observations, reducing the total number of pixels but hopefully retaining sufficient chemical signal to distinguish clusters. (3) Compute the distances after interpolating over the missing data. As the third approach is fast and the amount of missing data fairly minimal for these open clusters after our quality cuts, we use it for our test. We perform a linear spline interpolation in $T_{\rm eff}$ at each pixel across all stars to replace flux values flagged in the bitmask. However, option (3) is not reasonable for all samples or in all wavelength bands, and we anticipate that option (1) will be most useful when tagging larger samples of real observations.
	
After cuts to open cluster members with both good abundances and spectra, we are left with only three open clusters of sufficient size to test. The largest is NGC 6819, with 17 members in the OCCAM catalogue that survive our quality cuts. NGC 2682 and NGC 7789 each have 11 members, with the remaining clusters in the OCCAM catalogue having less than 10 members each. We insert the stars from the three largest open clusters into our set of stars from synthetic clusters.
	
Of the three open clusters, only NGC 6819 is recovered, and only then if the homogeneity threshold is lowered to 0.5, which implies that the group matched to NGC 6819 based on majority membership is nearly half full of members of other clusters. Further investigation reveals that while only one other cluster tends to contribute to the group that is matched to NGC 6819, it is still a major source of contamination.
	
The technique of extrapolating overall recovery fraction from the recovery fraction of open clusters holds promise for characterizing the success of DBSCAN in a way that is independent of knowledge of the true cluster membership. However, we need large homogeneous chemical surveys of open clusters to make this a robust statistic. While open clusters have been targeted extensively by spectroscopic surveys (e.g. Gaia-ESO - \citealt{Gilmore2012}) they should remain a focus in surveys that aim for significant Milky Way coverage on a single instrument.

\subsection{Comparison of chemical spaces}

Each of the chemical spaces we consider yields high homogeneity and completeness scores for the choices of $\tilde{\epsilon}$ that maximize the number of clusters recovered (Figure~\ref{fig:averageprop}). However, it is in abundance based chemical space that we achieve the highest recovery fraction; tagging the optimistic and theoretical abundances spaces recovers more than 30\% of the input clusters with high homogeneity and completeness. (Figure~\ref{fig:recv}).

While abundances may be too noisy with our conservative choice for intra-cluster chemical spreads to achieve a high recovery fraction, we expect better performance from the spectra. Spreads in abundances may be inflated by the post-processing required to derive abundances from the spectra, depending on the method used (e.g. \citealt{Holtzman2015}, \citealt{Nissen2018}), while the spectra are closer to the original observation, with less sources of noise to propagate. However, one challenge in constructing a chemical space with high resolution spectra is the high dimensionality of that space. When distances are computed between spectra, there are many pixels that do not inform about chemistry and only contribute noise or continuum information. Projecting spectra along their principal components reduces the amount of non-chemical information in the data, which gives a non-zero recovery fraction of about 0.1.

To assess the impact of the noise we add to the spectra, we gradually lower the spectra noise and track improvements in recovery fraction. The recovery fraction reaches a maximal value of about 0.5 even when noise in the spectra is assumed to be zero. This is a consequence of the fact that spectra of stars are generated based on different $T_{\rm eff}$ and $\log g$ for each star. Our approach to removing the influence of these properties on the spectrum (\S\ref{sec:synspec}) does not totally eliminate their effects, and we are left with non-chemical information in the spectra that prevents improvement of the recovery fraction. As expected, generating spectra for every star using the same $T_{\rm eff}$ and $\log g$ does allow the recovery fraction to climb to 1 for spectra-based chemical spaces, as in that case all variation between stars is due to chemistry. To obtain better recovery fractions using spectra directly, better methods to compute the distance between spectra will be necessary (e.g. \citealt{Itamar2019}). 

Distortion of abundances due to differences in $T_{\rm eff}$ and $\log g$ is not explicitly accounted for in our abundance space simulations. However, these properties  affect the absorption lines from which chemical abundances are derived, and so their bulk influence will be represented in the uncertainty in abundances we use to define the allowed chemical space spread in members of the same cluster. As the spread within a cluster decreases, the recovery fraction increases, mirroring the change in recovery fraction we observe when we reduce the contribution of noise pixels in the spectral space. As expected, the abundance space with the lowest chemical space spread within members of a cluster is the space that yields the highest recovery fraction. This further emphasizes the importance of improving the accuracy of and reducing abundance measurement uncertainty.

\subsection{Properties of successfully identified clusters}

We expect clusters that are successfully recovered by DBSCAN to share some characteristics: perhaps the largest clusters are found more reliably, or clusters more distinct from the centre of the chemical space distribution are easier for DBSCAN to group.

\begin{figure*}
\centering
\includegraphics[width = \linewidth]{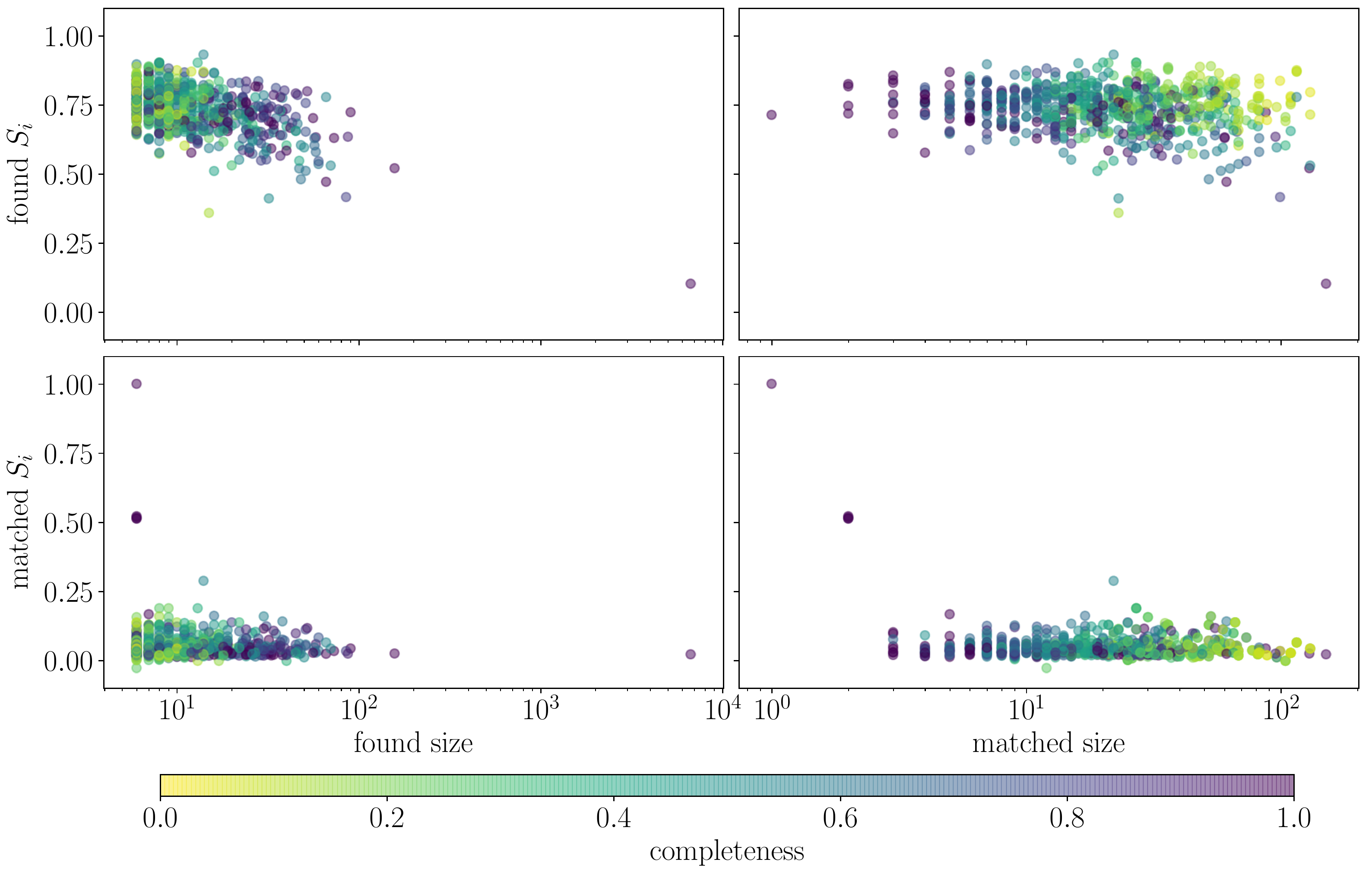}
\caption{The silhouette coefficient $S_i$ (equation~\eqref{eqn:silhouette}) as a function of the number of stars in the group (left column) and the number of stars in the cluster that the group was matched to (right column). The top row shows the $S_i$ computed for the groups found by DBSCAN, while the bottom row shows the $S_i$ for the true clusters that were matched to DBSCAN groups, where each $S_i$ computed relative to the full set of input clusters. At large matched size, there are both very high and very low completeness groups. The high completeness set is the set of large clusters dense enough that all members were placed in the same DBSCAN group. The low completeness set is a set of large clusters for which only their dense cores were correctly identified as groups by DBSCAN. This result is for a DBSCAN run on spectra projected into principal components, with $\tilde{\epsilon}=0.12$ and $N_{\rm pts} = 3$, limited to groups identified to have more than 5 members.}
\label{fig:sil}
\end{figure*}

To test the former hypothesis, we colour code a plot of silhouette coefficient versus group size and matched cluster size according to the completeness of the group (see Figure~\ref{fig:sil}). We show both the silhouette coefficient of the found group and of the cluster that was matched to it. We find a population of low completeness groups, marked in yellow, that have high found silhouette coefficient (top right panel), but low matched silhouette coefficient (bottom right panel). This indicates that DBSCAN can pick out the cores of large clusters, finding them as tightly bound groups but missing their outlying members. As a consequence of choosing each member's chemistry from a normal distribution in high-dimensional space, there are many more outlying members than one might expect based on intuition about a two-dimensional normal distribution, and so the completeness is driven down dramatically.

\begin{figure}
\centering
\includegraphics[width = \linewidth]{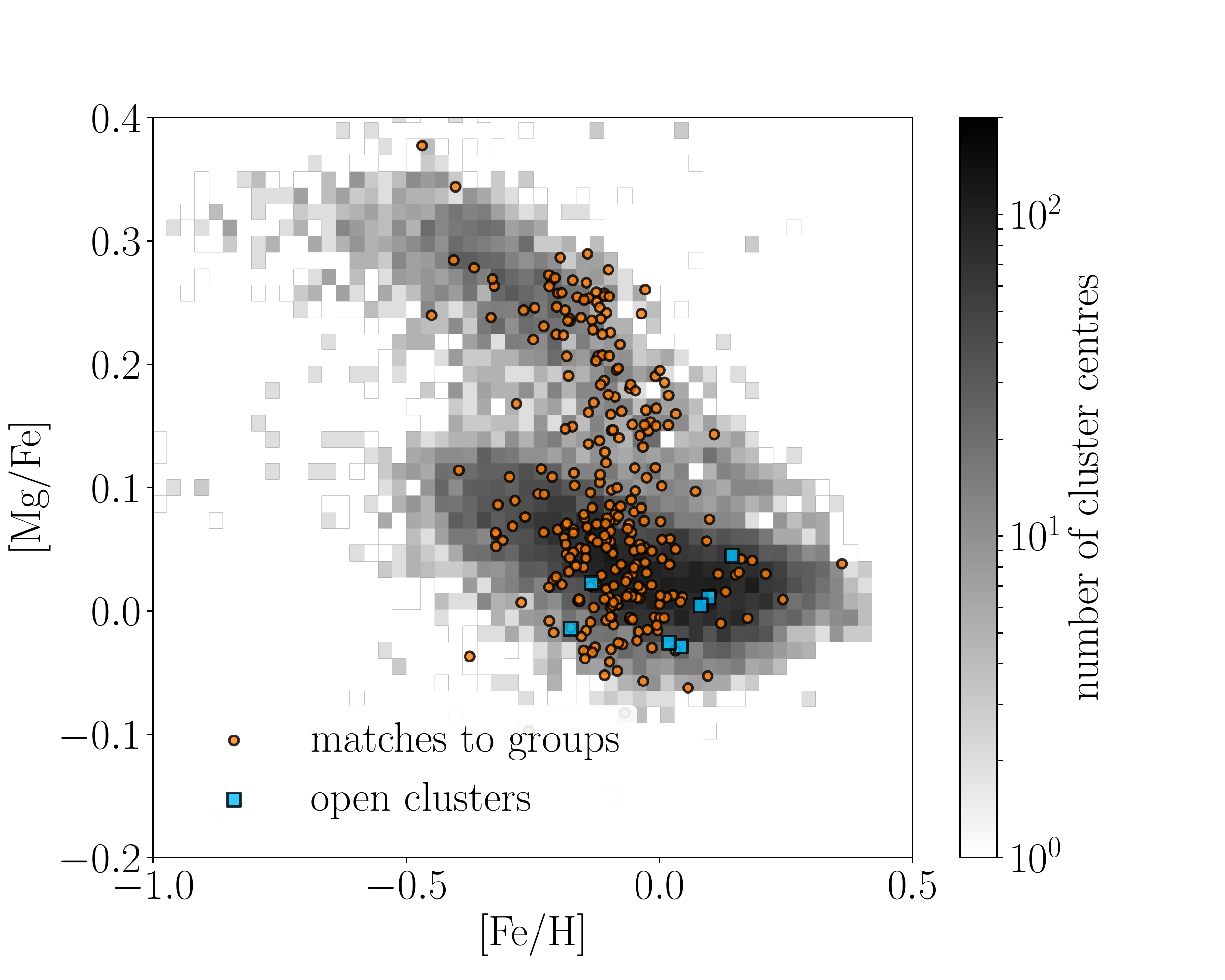}
\caption{One 2D projection of the 15 dimensional abundance chemical space, showing the abundances of Mg vs Fe. The distribution of true cluster centres is shown as the background histogram. Cluster centres matched to groups in DBSCAN (\emph{principal components}, $\tilde{\epsilon}=0.12$, $N_{\rm pts}=3$) with homogeneity and completeness greater than 0.7 are marked with orange circles. Blue squares mark the median abundances of cluster members of OCCAM open clusters. Note that the distribution of the matches does not trace the underlying distribution; there is an overabundance of matches at high [Mg/Fe]. We quantify this observation in Figure~\ref{fig:density}.}
\label{fig:chem}
\end{figure}

To better understand why some clusters are fully identified while others are limited to their cores, we consider the location of the clusters in chemical space. We show an example of 2D chemical space in Figure~\ref{fig:chem}, where we plot the abundance $\alpha$-element Mg relative to Fe against [Fe/H] for each cluster centre. The underlying distribution shown in a 2D histogram is the typical pattern for abundances of $\alpha$-elements in the Galaxy (e.g. \citealt{Hayden2015}). With orange circles, we plot the locations of clusters that are matched to groups in DBSCAN with high (>0.7) homogeneity and completeness. Blue squares mark the location of of open clusters from the OCCAM survey discussed in \S\ref{sec:ocrecv}.

\begin{figure}
\centering
\includegraphics[width = \linewidth]{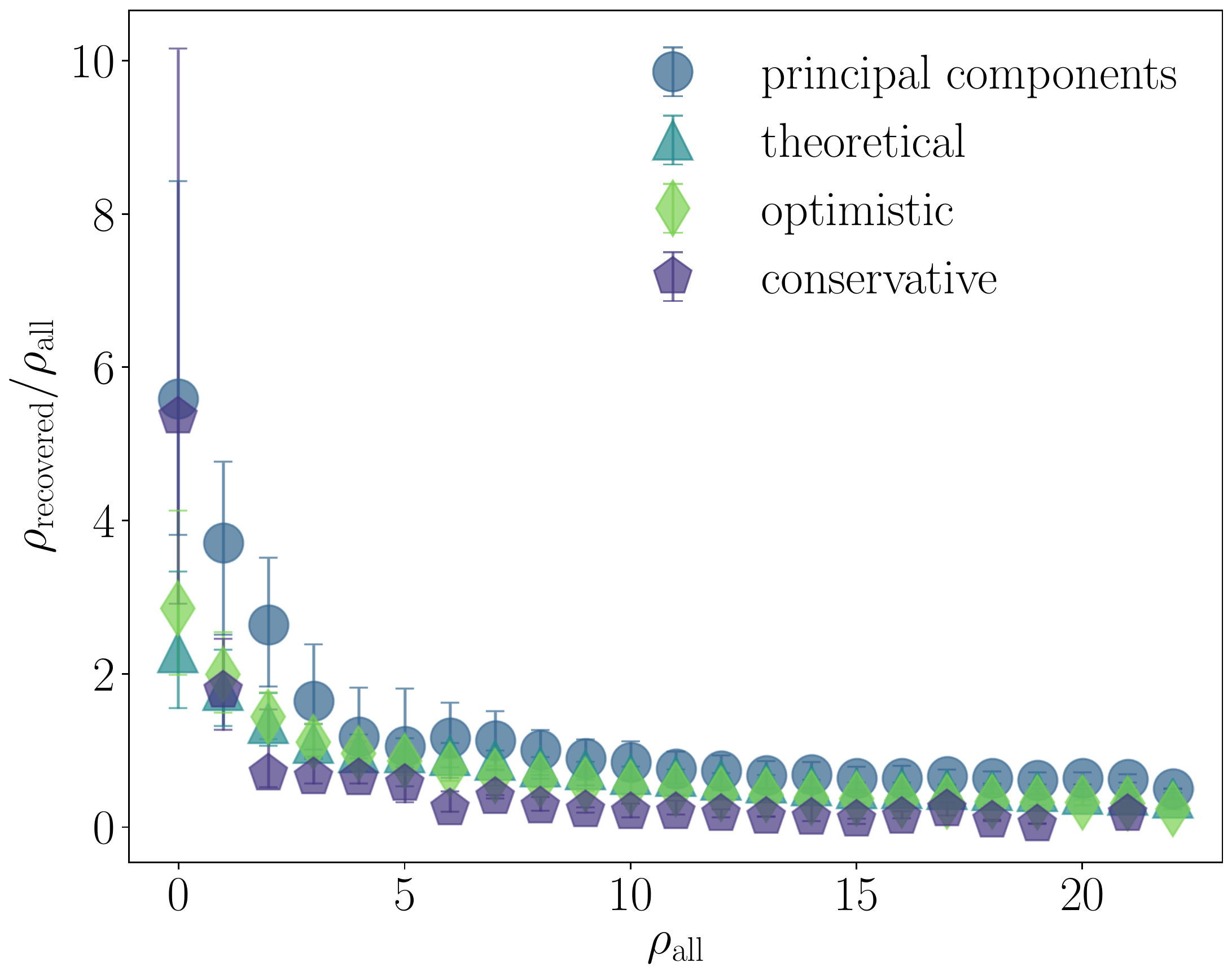}
\caption{The ratio of the density of group centres recovered by DBSCAN ($\rho_{\rm recovered}$) to the density of all cluster centres ($\rho_{\rm all}$), as a function of the density of all cluster centres. Density is computed in the two dimensional chemical space of [Mg/Fe] vs [Fe/H]. DBSCAN recovers more clusters where the total density of cluster centres was low for all types of chemical spaces.  Thus cluster finding is more efficient in low-density regions of abundance space.\label{fig:density}}
\end{figure}

The density of recovered clusters centres in Figure~\ref{fig:chem} does not trace the density of all cluster centres. Compared to the background, clusters are preferentially recovered at higher [Mg/Fe], where the underlying density of cluster centres is lower. We quantify this assessment by using a kernel density estimator (KDE) to model the density of all cluster centres ($\rho_{\rm all}$) and the density of recovered cluster centres ($\rho_{\rm recovered}$) in [Mg/Fe] vs [Fe/H] chemical space. The ratio of the two, $\rho_{\rm recovered}/\rho_{\rm all}$, is shown in Figure~\ref{fig:density} as a function of $\rho_{\rm all}$ for different chemical spaces. For all chemical spaces for which the recovery fraction was non zero, we find that $\rho_{\rm recovered}/\rho_{\rm all}$ is largest at small $\rho_{\rm all}$, indicating that DBSCAN has most success recovering clusters separated from their fellows in chemical space. While separations may seem small in our 2D projection, their distance from the bulk of cluster centres is more significant when all 15 abundance dimensions are considered.

\subsection{Modifying the simulated clusters}
\label{sec:modify}

Per the description in \S\ref{sec:creating}, we have three ways to modify the simulated clusters on which we run DBSCAN. We can change the following properties:
\begin{enumerate}
	\item the power law index of the cluster mass function, $\alpha$
	\item the volume of the annulus, $V$
	\item the number of stars observed, $N_{\rm survey}$
\end{enumerate} 

Parameters (i) and (ii) determine the total number of clusters included in the simulation. Specifying parameters (ii) and (iii) sets the sampling fraction, which governs how many member stars are observed from a given cluster. To explore how changes in the total number of clusters observed and the sampling fraction change the output of DBSCAN, we simulate a range of choices. Since our fiducial choices were $\alpha=2.1$ and $V=300\,\mathrm{kpc}^3$, we vary $\alpha$ between 0 and 2.6, and $V$ between 30 $\mathrm{kpc}^3$ and 1000 $\mathrm{kpc}^3$. 

As expected, the best choice for $\tilde{\epsilon}$ (the one that maximizes recovery fraction) remains consistent within each data type across all combinations of $V$ and $\alpha$. Our preliminary tests suggest that median completeness will decrease with decreasing $\alpha$, while median homogeneity should increase. As the CMF becomes flatter, there are more large clusters, and it is easier for them to dominate groups. But by the same token, it becomes more difficult to ensure all cluster members are in the same group when there are more of them per cluster. Median completeness should also correlate positively with survey volume; as the volume increases for a fixed $N_{\rm survey}$, fewer stars are sampled per cluster, making it less likely to sample an outlying star that would not be grouped with its fellows. 

Changes to the sampling rate might occur as part of differing survey plans choosing different numbers of stars to observe or accessing different portions of the Galaxy. However, our changes in volume also illustrate the impact of radial mixing on chemical tagging success. In our simulations, we assume that all members of each cluster remain inside the annulus where they are born, with a chance to be observed as they pass through the survey volume. In reality, clusters disperse after they form \citep{Lada2003}, and their members migrate away from their birth radius and height in the Galactic disk \citep{Sellwood2002}. This migration impacts the chemical structure of the Galaxy (e.g. \citealt{Schonrich2009}, \citealt{Hayden2015}) but will also impact our ability to recover birth clusters. When radial migration is significant, it moves stars into and out of the surveyed annulus. Moving stars out reduces the number available with which to detect a cluster, and moving stars into the region contributes to the background of stars who are the sole representative of their cluster. 

As radial migration grows stronger, a survey with fixed volume actually observes stars drawn from a larger effective volume, thanks to interlopers born outside that have migrated into the survey region. Thus our experimentation with changing volume emulates the effect various strengths of radial migration. Our largest possible volume of 1000 $\mathrm{kpc}^3$ is similar to that of the entire Milky Way's disk. Even in this extreme case, we recover clusters  with high homogeneity and completeness using our fiducial sample size.

\section{Discussion}
\label{sec:discussion}
 
The results we have presented in the previous section show that a density-based clustering technique can successfully recover input clusters when applied to a realistic dataset. We discuss in more detail some of our assumptions to create that dataset in the following section. In spite of these simplifying assumptions on the underlying data, we have explored a range of possible constraints on cluster homogeneity without making any requirements on cluster uniqueness. Since we choose random APOGEE stars to provide the central abundances for our clusters, no part of our process requires that the cluster centres be  more distinct from each other than current observations allow, and yet we enjoy remarkable success in recovering those clusters. Clusters are easiest to recover if they are in an area of chemical space with lower density, but can still be found in the presence of significant background. Our experiments varying survey volume in \S\ref{sec:modify} indicate that sufficiently large clusters are recovered well even when the sampling rate is decreased, which has the effect of increasing the background of single star clusters. This highlights one of DBSCAN's major advantages; the inclusion of the noise point classification makes it easy to discard any single-member clusters.

Our work with DBSCAN does have some limitations; we have worked primarily with a sample of only 50,000 stars. This is about a quarter of the size of APOGEE, and only a twentieth of the final GALAH survey. The 50,000 stars are presently treated democratically by DBSCAN, and while this is a viable approach when working with simulations, real data might necessitate some favouritism. Adding weighting to favour more trustworthy data would require modifying the process by which chemical space distances are computed. DBSCAN also does best when finding clusters of consistent density, and so multiple runs with different choices of $\tilde{\epsilon}$ may be needed to identify potential hierarchical structure.

Despite these potential challenges, our results have clearly demonstrated the remarkable power of DBSCAN to tease out structure that seems completely blended in two dimensional projection (e.g. Figure~\ref{fig:chem}). That the method does not require foreknowledge of the number of clusters in the sample is a great advantage over slightly faster techniques like k-means. A single DBSCAN run operating on our 50,000 star sample takes less than 30 seconds when pairwise distances between stars are precomputed, a process which only takes an additional minute. In addition, DBSCAN can identify non-spherical structure in chemical space - while this was not much exploited in our work with normally distributed clusters, it may prove an invaluable asset in real data. 

We outline below the primary assumptions of our simulation, with descriptions of how they might be relaxed in future work, followed by a discussion of the challenges to be overcome when applying DBSCAN to observations.

\subsection{Assumptions used to generate synthetic data }
\label{sec:assumptions}

The work presented here relies on several assumptions, which we describe below.

\subsubsection{Clusters are perfectly homogeneous}

The nature of chemical tagging is such that it assumes the presence of homogeneous clusters. We have taken that to an extreme limit here, assuming in our analysis that all differences between the chemistry of stars in the same birth clusters are due to measurement uncertainty. Especially in the case of the spectra we have failed to account for differences that might arise from variations  within the cloud in which the stars formed, such as pollution from massive stars that might go supernova before star formation is complete. This may be relevant in large star-forming clouds \citep{Bland-Hawthorn2010}. It would be useful to separate the noise into observational and intrinsic to examine how changes in either influence the resulting groups found by DBSCAN. However, DBSCAN recovers clusters well as long as they are at least as intrinsically homogeneous as our optimistic case for measurement uncertainties, and observed open clusters do seem to be that homogeneous \citep{Bovy2016}.

\subsubsection{Clusters have similar densities}

DBSCAN finds clusters according to a density defined by our choice of $\epsilon$ and $N_{\rm pts}$. This means it naturally finds clusters with density greater than $N_{\rm pts}/\epsilon^d$, where $d$ is the dimensionality of the chemical space. However, our method of cluster generation through normally distributing the chemical properties of member stars for each clusters means that larger clusters tend to appear less dense with respect to their neighbours (see Figure~\ref{fig:tsils}). One way to circumvent this would be a hierarchical approach, which would explore several possibilities for $\epsilon$ and use the resulting groups to merge together smaller groups (e.g. the hierarchical DBSCAN algorithm; \citealt{Campello2015}, \citealt{McInnes2017}). 

\subsubsection{Simplified mass in survey volume}

In our setup of the synthetic clusters described in \S\ref{sec:synthetic}, we make a few simplifying assumptions about the survey volume. We first assume a constant density for our Galaxy, which could be improved by incorporating a realistic density profile in the region of the Sun. We also assume that any members of clusters born in the survey annulus remain in the survey annulus. Future work could include the effects of radial migration moving cluster members into and out of the survey volume, which would have the effect of reducing the overall sampling rate by introducing more clusters in the same volume. Our preliminary experimentation in \S~\ref{sec:modify} indicates that this should overall increase completeness of recovered clusters, at the expense of lowering the homogeneity. 

We make no assumptions about what stars we could reasonably observe from a given cluster, using only the sampling rate to limit the number of observed members. A truly realistic simulation would incorporate the selection function of the survey it attempts to emulate, which requires setting not just chemistry for each star, but also stellar masses, ages, and distances from the observer. We are able to make our simulations relatively straightforward by ignoring observational preference for closer or brighter stars, but true surveys are not as random as our simulation.

\subsection{Applications to current and upcoming surveys}
\label{sec:applications}

In this work, we use the APOGEE survey as a template for the structure of our chemical spaces. More recent iterations of the ASPCAP now provide more than twenty abundances from APOGEE spectra \citep{Holtzman2018}, and surveys like GALAH measure even more abundances for each star. Incorporating additional abundances will give DBSCAN greater leverage to distinguish clusters; when we reduce the dimensionality of our abundance space to 5, overall homogeneity and completeness also declines. As long as additional abundances are not correlated with those already measured, the new information they provide will help distinguish clusters that might otherwise overlap.

However, switching from a survey like APOGEE to one like GALAH with an increased number of abundances will also impact the results of working with spectra. One potential difference in surveys not touched on in our exploratory analysis in \S\ref{sec:modify} is the result of changing the wavelength range in which we observe our stars. Different spectroscopic bands will be sensitive to different chemical species, offering alternative ways to distinguish stars. However, different bands also imply different targeting plans. APOGEE's focus on red giants is part of what made it a useful survey to emulate; while stars in the same birth cluster may have the same intrinsic abundances, it is the observed abundances that are important for chemical tagging. Stars in the same evolutionary phase are less likely to have surface abundance differences due to processes like atomic diffusion \citep{Dotter2017}, and so will lead to more successful chemical tagging. Even within a particular evolutionary phase, the contributions of some elements may be untrustworthy due to differences expected with different stellar mass (e.g. C and N; \citealt{Masseron2015}). Therefore, future surveys will need to balance the parallel goals of measuring chemistry for a sufficiently large sample of stars and ensuring those stars are close enough in evolutionary phase that they can be reasonably compared.

\section{Conclusions}
\label{sec:conclusions}
In this work, we have tested prospects for chemical tagging of spectroscopic observations with an efficient clustering algorithm, DBSCAN. We create synthetic chemical spaces that emulate observations, providing each star with fifteen chemical abundances and a continuum-normalized spectrum, ensuring that stars in the same cluster have the same chemical compositions, with normally distributed noise introduced to mimic observational uncertainties of varying magnitudes. We take the median abundances for each cluster as those of random red giant stars observed in APOGEE, so while we assume some varied levels of effective cluster homogeneity (depending on the amount of noise introduced in each star), we impose no constraint on the uniqueness of cluster chemical signatures. We define two categories chemical spaces; those based on spectra (with more than 7000 dimensions, each corresponding to a wavelength) and those based on the 15 derived abundances. We show that the DBSCAN algorithm can effectively recover input groups from these high-dimensional chemical spaces when they are populated by tens of thousands of stars. The density-based approach of this algorithm and its `noise star' classification allows it to ignore the chemical space background of stars which are the sole representative of their cluster in the survey. We use our multiple chemical spaces to compare the efficacy of using spectra with using derived abundances for clustering.

 While spectra offer more detailed chemical information than derived abundances, this information is diluted by the many continuum pixels that are not significant for chemically distinguishing stars. When using a Euclidean distance metric, these chemically insignificant pixels can overwhelm the differences between spectra introduced by variations in their intrinsic abundances. We reduce the influence of the continuum pixels by using PCA to identify the most significant pixels for distinguishing stars. By projecting spectra along their principal components, we achieve higher homogeneity in the found groups and recover 10 \% of the larger clusters from our input sample with homogeneity and completeness of at least 70\%. However, using abundance space with our `optimistic' requirements for effective cluster homogeneity \citep{Leung2019} increases the percentage of clusters recovered to 30\% with homogeneity and completeness of at least 70\%, indicating that decreasing our observational uncertainties in derived abundances may offer the most reliable approach to evaluating whether chemical tagging is viable in the Milky Way.
 
 These results are somewhat sensitive to the parameters of our simulated survey, and we perform tests of the success of chemical tagging with changing our cluster mass function, survey volume, and total number of stars observed. These suggest that chemical tagging should recover clusters at sampling rates even lower than that used for our fiducial sample, although this is sensitive to the underlying cluster mass function. Furthermore, our experimentation with changing simulation parameters suggests that significant radial migration is not likely to prevent the success of chemical tagging. Even when considering a sample where stars are drawn from a volume equivalent to that of the Milky Way's entire disk, DBSCAN was still able to recover clusters with high homogeneity and completeness.

Changing the survey volume modifies the sampling rate, allowing us to make predictions for different spectroscopic surveys. At our current simulation size limit, the sampling fraction is only $2\times 10^{-6}$, making the detection of clusters with fewer than $\sim$500,000 members unlikely. While we detect clusters smaller than this in our simulations, we do so only because we add a degree of randomness to the sampling rate for each cluster. Using the full APOGEE survey would increase that sampling fraction to $8\times 10^{-6}$ and thus reduce the minimum required members for detection to $\sim$125,000. Making use of the million spectra in the final GALAH set would bring that minimum requirement down to $\sim$25,000. Future surveys like the Milky Way Mapper (MWM) of SDSS V \citep{Kollmeier2017} will increase the number of spectroscopically surveyed stars to over 4 million. While this survey is designed to have lower signal to noise ratio than its predecessor, APOGEE, the MWM could lower the minimum number of members required to detect a cluster to 6250. Other planned surveys like the Mauna Kea Spectroscopic Explorer \citep{Zhang2016} and the 4-metre Multi-Object Spectroscopic Telescope \citep{deJong2016} will further increase the available sample of spectroscopically observed stars. This increased sampling rate will lead to more clusters sampled with more than 10 members. While only 535 clusters were sampled at that level in our fiducial simulation, our choice of cluster mass function indicates that we would expect to observe at least 10 members for more than 1,700 clusters in APOGEE and nearly 14,000 in GALAH, given our fiducial model for the Milky Way. If the MWM reduces abundances uncertainties from its target (close to our `conservative' case) to our `optimistic' case, its greater spatial coverage means it might sample 54,000 clusters with more than 10 members. Our predicted recovery fraction (Figure~\ref{fig:recv}) implies we might expect DBSCAN to find a tenth of these clusters as having more than ten members who are 70\% homogeneous when using a principal component reduction of the raw spectra. With sufficiently precise and accurate derived abundances (e.g. \citealt{Leung2019}), the percentage likely to be found with DBSCAN with this membership homogeneity rises above 30\%. The increase in chemical space resolution that comes of decreasing the uncertainty in a given star's abundances or spectrum is a crucial component of improving our ability to chemically tag birth clusters. This has been found in previous work with abundances (e.g. \citealt{Ting2015a}), but the importance of decreasing chemical space uncertainties cannot be overstated. 
 
 The true recovery fraction for observations from future surveys should be higher than the fraction we recover in our sample of 50,000 stars. As increasing the number of stars drives up the sampling rate, the overall homogeneity of the found groups should also increase. This increase will ensure that more clusters meet the homogeneity threshold they need to be considered 'recovered'. Thus these larger sample sizes and reduced chemical space uncertainties are essential for the birth cluster population studies that would allow us to learn the details of the star formation and chemical evolution history of our Galaxy.  

\section*{Acknowledgements}

We thank the anonymous reviewer for their careful reading of this work and for their thoughtful comments and suggestions.

NPJ is supported by an Alexander Graham Bell Canada Graduate Scholarship-Doctoral from the Natural Sciences and Engineering Research Council of Canada. NPJ and JB received support from the Natural Sciences and Engineering Research Council of Canada (NSERC; funding reference number RGPIN-2015-05235) and from an Ontario
Early Researcher Award (ER16-12-061).

Funding for SDSS-III has been provided by the Alfred P. Sloan Foundation, the Participating Institutions, the National Science Foundation, and the U.S. Department of Energy Office of Science. The SDSS-III web site is \url{www.sdss3.org}.

Funding for the Sloan Digital Sky Survey IV has been provided by the Alfred P. Sloan Foundation, the U.S. Department of Energy Office of Science, and the Participating Institutions. SDSS-IV acknowledges support and resources from the Center for High-Performance Computing at the University of Utah. The SDSS web site is \url{www.sdss.org}.

\bibliographystyle{mnras}
\bibliography{sim}
\label{lastpage}
\end{document}